\documentclass[11pt,a4paper]{article} 
       
        \usepackage[utf8]{inputenc}
        \usepackage{fontenc}
	    \usepackage[italian,english]{babel}
	    \usepackage{amsmath}
	    \usepackage{geometry}
	    \usepackage{amsthm}
	    \usepackage{mathrsfs}
	    \usepackage{bbold}
	    \usepackage{mathtools}
	    \usepackage{amsfonts}
	    \usepackage{graphicx}
	    \usepackage{setspace}
	    \usepackage{physics}
	    \usepackage{comment}
	    \usepackage{layout}
	    \usepackage{xcolor}
        \usepackage[big]{layaureo}
	    \usepackage{dsfont}	
	    \usepackage{slashed}
 	    \usepackage{amssymb}
	    \usepackage{tensor}
	    \usepackage{fancyhdr} 
 	    \usepackage{tikz-cd}
 	    \usepackage{braket}
 	    \usepackage{tikz}
 	    \usepackage[compat=1.1.0]{tikz-feynman}
 	    \usepackage{subfigure}
	    \usepackage{fancyhdr} 
	    \usepackage{booktabs}
	    \usepackage{braket}
	    \usepackage{caption}
	    \usepackage{wrapfig}

\usepackage{jheppub} 
\usepackage{natbib}
\bibstyle{JHEP}

\author[a]{Vladimir Bashmakov,} 
\author[b]{Nicola Gorini} 

\affiliation[a]{Department of Physics and Astronomy, Uppsala University, Box 516, SE-75120 Uppsala, Sweden}
\affiliation[b]{Dipartimento di Fisica, Universit\`a degli studi di Milano--Bicocca, and INFN, Sezione di Milano--Bicocca, Piazza della Scienza 3, I-20126 Milano, Italy }

\emailAdd{vladimir.bashmakov@physics.uu.se} 
\emailAdd{n.gorini1@campus.unimib.it}  

\abstract{We consider the IR phases of two-node quiver theories with $\mathcal{N}\,=\,1$ supersymmetry in $d\,=\,2\,+\,1$ dimensions. It turns out that the discussion splits into two main cases, depending on whether the Chern-Simons levels associated with the two nodes have the same sign, or the opposite signs, with the latter case being more non-trivial. The determination of the phase diagrams allows us to conjecture certain infrared dualities involving either two quiver theories, or a quiver and adjoint QCD. We also provide a short discussion on quivers possessing time reversal symmetry.}

	\newcommand{\beq}{\begin{equation}}
	\newcommand{\bea}{\begin{eqnarray}}
	\newcommand{\eea}{\end{eqnarray}}
	\newcommand{\eeq}{\end{equation}}

	\renewcommand{\a}{\alpha}
	\renewcommand{\b}{\beta}
	\newcommand{\g}{\gamma}

	\makeatletter
	\newcommand{\aextp}{\@ifnextchar^\@aextp{\@aextp^{\,}}}
	\def\@aextp^#1{\mathop{\bigwedge\nolimits^{\!#1}}}
	\makeatother
	
	\makeatletter
	\newcommand{\extp}{\@ifnextchar_\@extp{\@extp_{\,}}}
	\def\@extp_#1{\mathop{\aextp\nolimits_{\!#1}}}
	\makeatother

	\theoremstyle{definition}

\preprint{UUITP-45/21}

\title{Phases of $\mathcal{N}=1$ Quivers in $2+1$ Dimensions}

\begin{document}

\maketitle

\section{Introduction}

Three-dimensional quantum field theories with $\mathcal{N}=1$ supersymmetry\footnote{$\mathcal{N}=1$ supersymmetric theories in three dimensions have two real supercharges, forming a Majorana spinor.} constitute a remarkable bridge between theories with $\mathcal{N}=2$ supersymmetry, which are quite well understood thanks to holomorphy, and genuine non-supersymmetric theories, whose dynamics is a challenging subject. Given that we do not have neither non-renormalization theorems, nor localization techniques at our disposal, it may appear that $\mathcal{N}=1$ supersymmetry does not give any advantage compared to cases without supersymmetry. Nevertheless, in the recent years our understanding of these theories has overcome a new twist \cite{Gomis:2017ixy, Bashmakov:2018wts, Benini:2018umh, Gaiotto:2018yjh, Benini:2018bhk, Choi:2018ohn, Bashmakov:2018ghn, Aharony:2019mbc, Bashmakov:2019myq} (see also \cite{Gates:1983nr, Ohta:1997fr, Kitao:1998mf, Witten:1999ds, Bergman:1999na,Gremm:1999su, Acharya:2001dz,Schwarz:2004yj, Gukov:2002es, Forcella:2009jj, Amariti:2014ewa} for earlier considerations).

In particular, new tools for studying phase diagrams of $3d$ $\mathcal{N}=1$ theories were introduced and applied to the analysis of infrared (IR) dynamics of a vector multiplet coupled to matter in the adjoint or fundamental representations \cite{Bashmakov:2018ghn, Choi:2018ohn}. Among various interesting phenomena observed in those examples, we mention the existence of \textit{walls} in the parameter space where the Witten index jumps, as well as the presence of second-order (or higher) phase transition points. Both features seem to be ubiquitous for $3d$ $\mathcal{N}=1$ theories and stem from the fact that the theories possess real parameters.\footnote{$\mathcal{N}=2$ theories in three dimensions also have real parameters, which allow phase transition, but do not allow jumps of the Witten index \cite{Intriligator:2013lca}.}

In this paper we initiate the study of phase diagrams of $3d$ $\mathcal{N}=1$ theories with the gauge group given by a product of several non-Abelian factors and bi-fundamental matter coupled to it, known as \textit{quiver} gauge theories. Such theories may arise as worldvolume theories of BPS domain walls and interfaces in four-dimensional supersymmetric theories \cite{Acharya:2001dz, Gaiotto:2017tne, Bashmakov:2018ghn}, as well as world-volume theories of M2/D2 branes on backgrounds with two preserved supercharges (e.g. M2 branes probing $Spin(7)$ cones \cite{Forcella:2009jj}). There exists extensive literature on $\mathcal{N}=2$ quiver theories (for a recent discussion see e.g. \cite{Benini:2011mf, Amariti:2014lla,Benvenuti:2016wet,Amariti:2017gsm,Benvenuti:2017kud,Benvenuti:2017bpg,Nedelin:2017nsb,Aprile:2018oau,Amariti:2019pky,Pasquetti:2019uop, Pasquetti:2019tix,Jain:2019lqb, Benvenuti:2020gvy,Benvenuti:2020wpc} and references therein) and also some results regarding non-supersymmetric quiver theories \cite{Karch:2016aux, Jensen:2017dso, Aitken:2019mtq}. For prior studies of $\mathcal{N}=1$ quivers instead see \cite{Amariti:2014ewa}.

In the present work we restrict ourselves to the two-node quivers with $SU(2)$ or $U(2)$ gauge groups and with one bi-fundamental matter multiplet. This simple setup allows for a detailed treatment and is still rich enough to accommodate a variety of interesting phenomena. In particular, we reveal a rather non-trivial phase structure of these theories and identify a collection of $\mathcal{N}=1$ superconformal field theories (SCFT). Basing on our understanding of phase diagrams, we are able to conjecture certain infrared dualities, some of which can be understood as the dualization of a node of a quiver, and some others as the confinement of a node. Finally, when the gauge groups of the two-node quiver are the same, and the Chern-Simons levels are chosen to be opposite, the theory turns out to be time-reversal invariant. This symmetry gives rise to certain non-renormalization theorems \cite{Gaiotto:2018yjh} which put strong restrictions on the form of the effective superpotential.

To get some preliminary ideas about the features of the phases of quiver theories we are going to study, let us consider the $SU(2)_{k_1}\times SU(2)_{k_2}$ theory with a bi-fundamental multiplet $\Phi$ (we assume for a moment that $k_1,k_2>0$). This theory has two discrete dimensionless parameters, the Chern-Simons levels $k_1$ and $k_2$, and three continuous dimensionful parameters, the two gauge couplings $g_1,g_2$ and the matter mass $m$. We can construct two independent continuous dimensionless parameters\footnote{More explicitly, one of the parameters is going to be $\frac{g_1}{g_2}$, and the other is a ratio of the mass and some combination of the couplings, \textit{e.g.} $\frac{m}{g_1^2+g_2^2}$.}, and therefore generically we expect to find a two-dimensional phase space.

In the large mass limit, $|m|\gg g_{1,2}^2$, matter can be integrated out semiclassically. If $m>0$, we get at low energies
\begin{equation}
    \mathcal{N}=1\qquad SU(2)_{k_1+1}\times SU(2)_{k_2+1},
\end{equation}
where the one-loop renormalization of Chern-Simons levels is taken into account (see Section \ref{Review} for some details). This theory further flows to a purely topological CS theory \cite{Witten:1999ds},
\begin{equation}
    SU(2)_{k_1}\times SU(2)_{k_2}\quad\text{TQFT}.
\end{equation}
If instead $m<0$, we get 
\begin{equation}\label{TwoorOneGoldstino}
    \mathcal{N}=1\qquad SU(2)_{k_1-1}\times SU(2)_{k_2-1},
\end{equation}
which can either flow to a CS theory, or break SUSY and give rise to a Goldstino \footnote{Eq. \ref{TwoorOneGoldstino} may suggest that when both vector multiplets break supersymmetry, two Goldstini arise. A more natural picture is that we get one massless Majorana fermion (the Goldstino) and one very light Majorana fermion, with the mass suppressed by the scale of $m$.} with a decoupled topological sector. We note in particular that the Witten indices differ in these two limits, which indicates the existence of a wall that separates the two large mass limits (see Section \ref{Review} for some details).

In order to understand the transition between the two asymptotic phases, it is useful to consider the behaviour of the theory near the line (one dimensionless parameter involving the mass is fixed while the other is left free) where the matter is massless and the theory has a classical moduli space of vacua. In fact, it is natural to assume that new vacua appear near this line to compensate the jump of the Witten index; this is going to be the wall mentioned above.

To describe more explicitly the classical moduli space, let us note that every complex matrix $M$ possesses a \textit{singular value decomposition} of the form $M\,=\,U\,\Sigma\,V^{\dagger}$, where $U$ and $V$ are unitary, and $\Sigma$ is diagonal with real non-negative entries (the matrices can be chosen such that the diagonal entries appear in descending order). Thus, we can apply two gauge transformations, one for each factor of the gauge group, to put the matrix $\Phi$ in a diagonal form with real diagonal entries all multiplied by an overall phase. In total, the classical moduli space for the $SU(2)\times SU(2)$ case is given by
\begin{equation}\label{ClassicalModuliSpaceSU2}
 \mathcal M = \mathbb S^1\,\times\,\mathbb{R}^2\,/\,S_2.
\end{equation}
whereas for the $SU(2)\times U(2)$ case it is instead given by
\begin{equation}\label{ClassicalModuliSpaceU2}
 \mathcal M = \mathbb{R}^2\,/\,S_2.
\end{equation}
The effective superpotential is expected to receive radiative corrections, and the classical moduli space is lifted at the quantum level.\\\\
The rest of the paper is organised in the following way: In Section \ref{Review} we review some specific aspects of $\mathcal N=1$ gauge theories with different matter contents, which are essential for the rest of the paper, including the IR behaviour of a pure $\mathcal{N}=1$ vector multiplet, a vector multiplet coupled to matter in the fundamental representation, and a vector multiplet coupled to matter in the adjoint representation. In Section \ref{Superpotential} we compute the effective superpotential for the $SU(2)\times SU(2)$ class of quivers and study its behaviour in some physically relevant asymptotic limits. This result will allow us to study the phase diagrams of these models which we will present in Section \ref{SUSU}. We then proceed with the study of the phase diagrams for $SU(2)\times U(2)$ models in Section \ref{SUU}. For this case we do not have the explicit effective potentials, so we try to do some reasonably looking assumptions about their vacuum structure close to the wall. In Section \ref{Dualities} we propose certain dualities involving quiver theories, as well as the adjoint SQCD. In section \ref{TimeReversalInvariantModels} we make some comments regarding the dynamics of quiver theories enjoying time reversal invariance. Section \ref{Outlook} contains conclusions and future directions.

\section{Review of previous results}\label{Review}

\subsection{Chern-Simons levels}

Since we are going to deal with Chern-Simons terms, we start by stating the related conventions. In three dimensions, the Yang-Mills theory admits a gauge invariant deformation given by the Chern-Simons term. This is \textit{a priori} characterized by its level $k_{bare}\in\mathbb{Z}$. Instead of characterizing it directly by $k_{bare}$, we prefer to follow the common practice and introduce the level $k$ related to the bare one by 
\begin{equation}\label{UConsistency}
    k\,=\,k_{bare}-\tfrac{1}{2} \sum_f T(R),
\end{equation}
where the sum is taken over fermions in the representation $R$ of the given gauge group and $T(R)$ is the Dynkin index\footnote{We recall to the reader that for $SU(N)$ we have $T(F)=\frac{1}{2}$ and $T(A)=\frac{N}{2}$ for respectively the fundamental and the adjoint representations.} of the real representation $R$. Notice that the level $k$ defined above transforms as $k\rightarrow-k$ under time-reversal symmetry.

When a Majorana fermion of mass $m$ in a real representation $R$ is integrated out, the level gets shifted as follows\footnote{In the case of fermions in a complex or a pseudo-real representation, the $\tfrac{1}{2}$ factor has to be dropped from both equations \eqref{UConsistency} and \eqref{massConsistency}.}
\begin{equation}\label{massConsistency}
    k\rightarrow k+\tfrac{1}{2}\text{sign}(m)T(R).
\end{equation}

When all fermions are integrated out, the resulting level is sometimes referred to as the \textit{infrared level}. Gauge invariance requires that the bare level $k_{bare}$ and the infrared level $k_{IR}$ are integrally quantized. At the same time, the level $k$ used to label a theory may be either integrally quantized or half-integrally quantized, depending on the number of fermions in the theory and their representations (charges).

\subsection{Level-rank duality}
Theories with $U(N)$ gauge groups may have two Chern-Simons terms, the single trace and the double-trace ones. According to the standard definition, we have
\begin{equation}
    U(N)_{k_1,k_2}\,=\,\frac{SU(N)_{k_1}\times U(1)_{N k_2}}{\mathbb{Z}_N},
\end{equation}
and consistency requires that $k_2=k_1\ \text{mod}\ N$. With $C$ being a $U(N)$ gauge field,
the Chern-Simons Lagrangian is
\begin{equation}
    \frac{k_1}{4\pi}\text{Tr}\left(CdC-\frac{2i}{3}C^3 \right)+\frac{k_2-k_1}{4\pi N}(\text{Tr}C)d(\text{Tr}C).
\end{equation}

There are two important facts about Chern-Simons TQFTs to be mentioned. First, it happens sometimes that different Chern-Simons theories may in fact describe the same TQFT. An important example of this phenomenon, used in what follows, is given by Level-Rank duality (see \cite{Hsin:2016blu} for a modern discussion):
\begin{eqnarray}
    \label{LevelRankSUU}
    &U(N)_{\pm k,\pm k}\longleftrightarrow SU(k)_{\mp N},\\
    \label{LevelRankUU}
    &U(N)_{k,k\pm N}\longleftrightarrow U(k)_{-N,-N\mp k},
\end{eqnarray}
Second, we note that while in general time reversal transformation maps a TQFT to a different TQFT, there are some which turn out to be $T$-invariant. The duality \eqref{LevelRankUU} implies that $U(N)_{N,2N}$ is of this kind. 

\subsection{$SU(N)_k$ Vector multiplet}\label{ReviewVectorMultiplet}
$\mathcal{N}=1$ $SU(N)_k$ vector multiplet, consisting of a gauge field $A$ and a Majorana fermion $\lambda$ in the adjoint representation, was first studied in \cite{Witten:1999ds}. For $k\geq N/2$ the theory flows to the $SU(N)_{k-\frac{N}{2}}$ Chern-Simons theory. For large values of the level, i.e. $k\gg1$, this can be understood semiclassically by integrating out the gaugino, whose mass is $m_{gaugino}=-\frac{k_{bare} g^2}{2\pi}$. This gives the $-\frac{N}{2}$ shift of the Chern-Simons level. \\
In the resulting Yang-Mills-Chern-Simons theory, the propagating modes are massive \cite{Deser:1981wh,Deser:1982vy} and at low energies only topological degrees of freedom survive. The Witten index in this regime is given (up to an overall sign\footnote{As a general comment, the sign depends on the number of Majorana fermions with negative mass in the theory \cite{Witten:1982df}, which we can call $n_F$. The overall sign is then given by $(-1)^{n_F}$}.) by
\begin{equation}
    I\,=\,\frac{1}{(N-1)!}\left(k\,-\,\frac{N}{2}\,+\,1 \right)\left(k\,-\,\frac{N}{2}\,+\,2 \right)...\left(k\,+\,\frac{N}{2}\,-\,1 \right),
\end{equation}
which is just the partition function of the low-energy Chern-Simons TQFT on the torus.\footnote{Equivalently, it can be computed by counting the number of inequivalent Wilson lines of the theory.} 

For $0\leq k<N/2$ the Witten index vanishes, and it was conjectured in \cite{Witten:1999ds} that supersymmetry is spontaneously broken in such models, thus a massless Goldstino emerges in the IR limit. This conjecture found significant support in the results of \cite{Gomis:2017ixy}, where it was also claimed that, in addition to the Goldstino $G_{\alpha}$, there must be topological degrees of freedom described by a $U\left(\frac{N}{2}-k\right)_{\frac{N}{2}+k,N}$ Chern-Simons theory. \footnote{The fact that the Goldstino alone is not enough to describe the IR physics follows from the following observations: at any value of the level $k$ there is a 1-form symmetry in the UV, moreover, for $k=0$ the UV theory enjoys the time reversal symmetry. There are 't Hooft anomalies associated with both symmetries that cannot be matched by the Goldstino along, and some other d.o.f. are required.} This TQFT has a non-perturbative origin, and its precise form follows from certain dualities.

\vskip 30pt

\subsection{Vector multiplet coupled to an adjoined matter multiplet}
Matter multiplet consists of a real scalar and a Majorana fermion, and $\mathcal{N}=1$ supersymmetry allows real superpotential for matter; in what follows we will be interested in the tree-level superpotentials given only by the mass term. The dynamics of an $SU(N)$ vector multiplet coupled to a matter multiplet in the adjoint representation as a function of the matter mass parameter was described in details in \cite{Bashmakov:2018wts}. Here we review the case of $SU(2)$ gauge group, relevant for the rest of the paper.

For $k\geq2$ and $m\gg 0$, we can integrate out the matter multiplet and get the pure $SU(2)_{k+1}$ vector multiplet. In this limit the theory flows to a supersymmetric vacuum hosting the $SU(2)_k$ CS theory. When we consider $m\ll0$ instead, we get the pure $SU(2)_{k-1}$ vector multiplet, which flows to a supersymmetric vacuum with  $SU(2)_{k-2}$ CS theory. Thus, there are two asymptotic phases with different Witten indices. The transition between these two phases happens in two stages. At $m=0$ the effective potential develops a flat direction, and for small and positive mass a new supersymmetric vacuum supporting the $U(1)_{2k}$ CS theory comes in from infinity of the field space: this new vacuum compensates the jump of the Witten index. At some finite mass value two vacua merge through a second order phase transition and produce a single vacuum visible at large positive masses.

For $k=1$ and large positive mass the picture is not much different from the previous case. Indeed, we get a single supersymmetric vacuum supporting the $SU(2)_1$ TQFT. On the contrary, when the mass is large and negative, by integrating out matter we get the $SU(2)_0$ vector multiplet which breaks supersymmetry. The theory in the IR limit is then given by\footnote{We warn the reader that the $U(1)_2$ here is of the origin discussed at the end of Section \ref{ReviewVectorMultiplet}.}
\begin{equation}\label{u12vacuum}
    G_{\alpha}\,+\,U(1)_2.
\end{equation}
At $m=0$ the effective potential develops the flat direction, and for small positive mass we find one supersymmetric vacuum with $U(1)_2$ TQFT, which came in from infinity of the field space. Using the fact that $U(1)_2$ is time-reversal invariant and applying the Level-Rank duality (cf. \cite{Hsin:2016blu}), 
\begin{equation}
U(1)_2\longleftrightarrow U(1)_{-2}\overset{\text{L-R}}{\longleftrightarrow} SU(2)_1,
\end{equation}
we recognise the same vacuum as we have seen at the large positive mass. Therefore, we do not have a phase transition in this case. We note that at positive and small mass values, the non-supersymmetric vacuum coexists with the supersymmetric one, and so the former is meta-stable.

Finally, for $k=0$, in the $m=0$ point we have the supersymmetry enhanced to $\mathcal{N}=2$, and the theory shows a runaway behaviour \cite{Affleck:1982as}. For non-zero masses the runaway behaviour stabilizes and the superpotential shows a single trivial supersymmetric vacuum for both positive and negative masses. 

\subsection{Vector multiplet coupled to fundamental matter multiplets}
In this section we review the physics of $SU(2)$ and $U(2)$ vector multiplets coupled to $F$ matter multiplets in the fundamental representation as a function of the matter mass $m$ (we assume that all flavours are given the same mass). This is a particular case of the results in \cite{Choi:2018ohn}.

Consider first $U(2)_{k+1,k}$ theories with $F$ fundamentals.
\begin{itemize}
\item \textbf{$F=1$:}  At large positive mass, by integrating out the matter, we get a single vacuum with $U(2)_{k+\frac{3}{2},k+\frac{1}{2}}$ vector multiplet. For large and negative mass we get a single vacuum with $U(2)_{k+\frac{1}{2},k-\frac{1}{2}}$ vector multiplet. In the intermediate phase $0<m<m_*$ (where $m_*$ is some positive value) in addition to the vacuum seen at negative masses, a new vacuum with $U(1)_{k+\tfrac{1}{2}}$ vector multiplet appears. At $m=0$ there is a wall, and $m_*$ is a phase transition point, where the two vacua of the intermediate phase merge, and produce the large positive mass vacuum.

\item \textbf{$F\geq2$:}  At large positive masses, we get a single vacuum with $U(2)_{k+\tfrac{2+F}{2},k+\tfrac{F}{2}}$ vector multiplet. For large and negative masses we get a single vacuum with $U(2)_{k+\tfrac{2-F}{2},k-\tfrac{F}{2}}$ vector multiplet. In the intermediate phase $0<m<m_*$ in addition to the negative mass vacuum, two new vacua appear. One of them hosts the $\frac{U(F)}{U(F-1)\times U(1)}$ $\mathcal{N}=1$ non-linear sigma model (coming from spontaneously broken $SU(F)$ global symmetry) together with the decoupled $U(1)_k$ vector multiplet. The second new vacuum hosts the $\frac{U(F)}{U(F-2)\times U(2)}$ $\mathcal{N}=1$ NLSM. Again, at $m=0$ there is a wall, and at $m=m_*$ is a phase transition point.
\end{itemize}
A peculiarity of this type of models is that, in contrast to expectations, the three vacua living in the intermediate phase merge in just a single phase transition instead of two distinct ones. This phenomenon was dubbed \textit{supercriticality}.

\vskip 30pt

Next we discuss $SU(2)_{k+\tfrac{F}{2}+1}$ theories with $F$ fundamentals. It worth reminding that the global symmetry here is not just $U(F)$, but rather $Sp(F)$, thanks to the pseudo-real nature of the $SU(2)$ fundamental representation. For large and positive mass we get a single vacuum with $SU(2)_{k+F+1}$ vector multiplet, while for negative masses  we have $SU(2)_{k+1}$ vector multiplet. In the intermediate phase with $0<m<m_*$, in addition to the negative mass vacuum, a new vacuum come in from infinity and supports the $\frac{Sp(F)}{Sp(1)\times Sp(F-1)}$ $\mathcal{N}=1$ NLSM.

The authors of \cite{Choi:2018ohn} described a family of dualities between such SQCD theories, among which there is the following \footnote{V.B. thanks Adar Sharon for a discussion of this duality.} 
\begin{equation}\label{ParentDuality}
    U(2)_{2,1}+2\Phi\longleftrightarrow SU(2)_{-2}+2\tilde{\Phi}.
\end{equation}
This duality is quite subtle and, if correct, has quite far-reaching consequences, such as symmetry enhancement on the $U(2)$ side from $SU(2)\times U(1)$ to $Sp(2)$ at the IR fixed point. A puzzling aspect about this duality is the mismatch of the phases (described above), when the theories are deformed away from the fixed point by the mass deformation. In particular, while the $U(2)$ theory has three vacua in the intermediate phase, the the $SU(2)$ theory has just two of them. A possible explanation of this apparent mismatch is the breaking of supercriticality argumentation due to \textit{e.g.} non-perturbative corrections. Giving up supercriticality, one can imagine the intermediate phase with two vacua, matching exactly the $SU(2)$ vacua.
\vspace{0pt}
\begin{center}
$ {\ast}\,{\ast}\,{\ast} $
\end{center}
In all the examples above, while one can  easily get the vacua for large mass values, it is more subtle to understand the dynamics in the intermediate phase, for which one needs to know the effective superpotential. For this reason, coming back to our original problem of interest, we turn to the evaluation of the effective superpotential for the two-node quiver theories as described in the introduction.

\section{Effective superpotential for $SU(2)\times SU(2)$ theories}\label{Superpotential}

In this section we consider two-node quiver theories with $SU(2)_{k_1}\times SU(2)_{k_2}$ gauge group and with one bi-fundamental matter multiplet and compute the effective superpotential, the superspace analogue of the effective potential discussed first by Coleman and Weinberg \cite{Coleman:1973jx}. It turns out to be useful to use supergraph formalism for this computation, in terms of which the leading contribution comes at 1-loop order.\footnote{One can work in components as well, but then the two-loop computation would be needed, since the one-loop result vanishes.} We follow closely the discussion of \cite{Choi:2018ohn}, a detailed discussion of the $\mathcal{N}=1$ superspace and supergraph formalism can be found in \cite{Gates:1983nr}. In fact, up to some point, the discussion can be held at a more general level, therefore we will start by considering the case of $G_1\times G_2$ gauge groups and then specialize to the case of $SU(2)\times SU(2)$.

\begin{figure}
    \centering
    \includegraphics[scale=0.4]{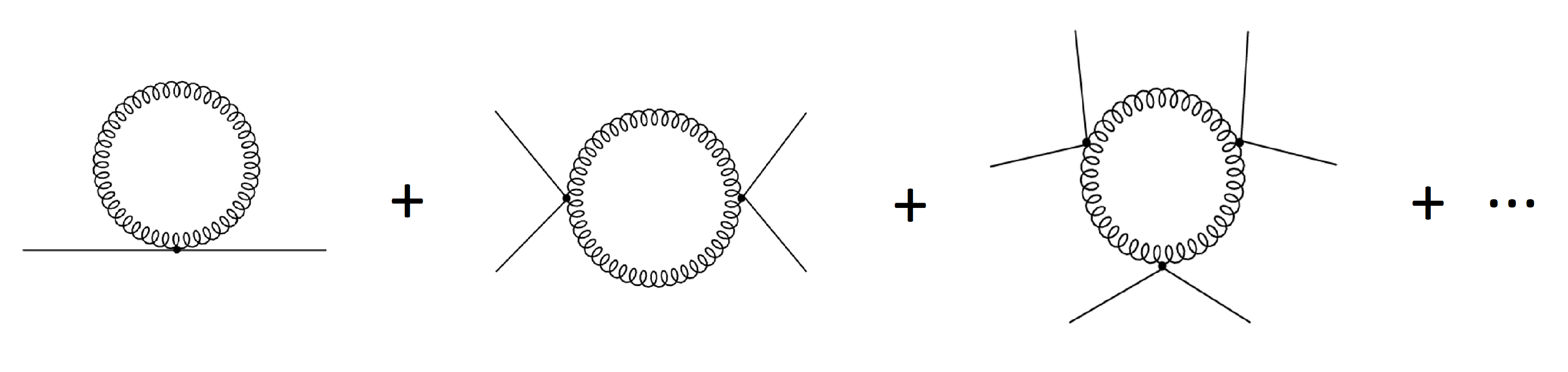}
    \caption{Feynman Supergraphs contributing to the 1-loop effective Superpotential.}
    \label{fig:supergraphs}
\end{figure}

The effective superpotential is of the form \cite{Choi:2018ohn}
\begin{equation}\label{superpot}
    \mathcal{W}_{\text{eff}}\,=\,\int\frac{d^3p}{(2\pi)^3}d^2\theta'\,\delta(\theta-\theta')\Sigma(p,\theta')\delta(\theta'-\theta),
\end{equation}
where $\Sigma$ at the one-loop order is given by the sum of the diagrams in Figure \ref{fig:supergraphs}.\\
We note that the form of the series expansion is exactly the same as can be found in \cite{Coleman:1973jx}, albeit with Feynman rules replaced by their superspace counterpart. We thus state the relevant superspace Feynman rules now.

Let $\Gamma_\alpha^A,\,\hat\Gamma_\alpha^M$ be vector multiplet superfields of the first and the second gauge groups, respectively; here $A, M$ are colour indices and $\alpha$ is the spinor index. The two gauge propagators are (\ref{gaugeprop})
\begin{align}
\braket{\Gamma_\alpha^A(p)\Gamma^{B,\beta}(-p)}&=\delta^{AB}\ \frac{\delta_{\alpha}^{\beta}(\kappa_1\,D^2\,+\,p^2)\,+\,(\kappa_1\,-\,D^2)\,{p_{\alpha}}^{\beta}}{p^2\,(\kappa_1^2\,+\,p^2)}\equiv \delta^{AB}{(\Delta_1)_{\alpha}}^{\beta},\\
\braket{\hat\Gamma_\alpha^M(p)\hat \Gamma^{N,\beta}(-p)}&=\delta^{MN}\ \frac{\delta_{\alpha}^{\beta}(\kappa_2\,D^2\,+\,p^2)\,+\,(\kappa_2\,-\,D^2)\,{p_{\alpha}}^{\beta}}{p^2\,(\kappa_2^2\,+\,p^2)}\equiv \delta^{MN}{(\Delta_2)_{\alpha}}^{\beta},
\end{align}
where $\kappa_i=\frac{k g_i^2}{2\pi}$. \\
There are also three types of vertices joining two gauge fields and two matter fields\footnote{As usually, there also exist a cubic vertex with two matter legs and one gauge boson leg but they do not contribute since, in the Landau gauge, the gauge propagators are transverse, namely $(\Delta_{1,2})_{\alpha}^{\ \beta} D_{\beta}=0$.}:\\
we get the rules 
\begin{align}
\Gamma^{(4)}_{\Gamma\Gamma\bar\phi\phi}&=\braket{\Gamma_\alpha^A(p)\Gamma^{B,\beta}(q){\Phi_{i}}^{\hat{j}}(r){\bar{\Phi}_{\hat{k}}}^{\ l}(-p-q-r)}\,=\,-\frac{g_1^2}{2} (T^{(A})_i^{\ n}(T^{B)})_n^{\ l}\,{\delta^{\hat{j}}}_{\hat{k}}\,{\delta_{\alpha}}^{\beta},\\
\Gamma^{(4)}_{\hat\Gamma\hat\Gamma\bar\phi\phi}&= \braket{\hat{\Gamma}_\alpha^M(p)\hat{\Gamma}^{N,\beta}(q){\Phi_{i}}^{\hat{j}}(r){\bar{\Phi}_{\hat{k}}}^{\ l}(-p-q-r)}\,=\,-\frac{g_2^2}{2} (K^{(M})_{\hat{k}}^{\ \hat{n}}(K^{N)})_{\hat{n}}^{\ \hat{j}}\,{\delta_{i}}^{\l}\,{\delta_{\alpha}}^{\beta},\\
\Gamma^{(4)}_{\Gamma\hat\Gamma\bar\phi\phi}&=\braket{\Gamma_\alpha^A(p)\hat{\Gamma}^{M,\beta}(q){\Phi_{i}}^{\hat{j}}(r){\bar{\Phi}_{\hat{k}}}^{\ l}(-p-q-r)}=\frac{1}{2} g_1g_2 (T^A)_i^{\ l}(K^M)_{\hat{k}}^{\ \hat{j}}\,{\delta_{\alpha}}^{\beta}.
\end{align}
Above unhatted lower case letters $i,j,...$ are fundamental indices and $A,B$ are adjoint indices of $G_1$, with $T^A$ being its generators. Similarly, hatted lower case letters are fundamental indices and $M,N$ are adjoint indices of $G_2$, with $K^A$ being its generators.

We can now proceed with the computation of $\Sigma$. For this purpose, we find it useful to introduce the following matrices
\begin{subequations}\label{matrices}
\begin{align}
    M^{AB}&=g^2_1\text{Tr}\,\bar{\Phi}T^{(A} T^{B)}\Phi,\\
    N^{MN}&=g^2_2\text{Tr}\,\Phi K^{(M} K^{N)}\bar{\Phi},\\ 
    G^{AM}&=-g_1g_2\text{Tr}\,\bar{\Phi}T^A\Phi K^M,
\end{align} 
\end{subequations}
which allow us a more compact treatment of the computation, and in terms of which the expansion of Figure \ref{fig:supergraphs} takes the form 
\begin{equation}\label{sigma}
\begin{aligned}
\Sigma_{\text{1-loop}}&=-\tfrac{1}{2}\,\text{Tr}\,M\,\Delta_1\,-\,\tfrac{1}{2}\,\text{Tr}\,N\,\Delta_2\,+\\
&\ \quad +\tfrac{1}{4} \text{Tr}\,(M\,\Delta_1)^2\,+\,\tfrac{1}{2} \text{Tr}\,(G)^\dagger\,\Delta_1\,G\,\Delta_2+\,\tfrac{1}{4} \text{Tr}\,(N\,\Delta_2)^2\,-\\
&\ \quad -\tfrac{1}{6} \text{Tr}\,(M\,\Delta_1)^3\,-\,\tfrac{1}{2} \text{Tr}\,G^\dagger\,\Delta_1\,M\,\Delta_1\,G\,\Delta_2\,-\,\tfrac{1}{2} \text{Tr}\,G\,\Delta_2\,N\,\Delta_2\,G^\dagger\,\Delta_1\,-\,\tfrac{1}{6} \text{Tr}\,(N\,\Delta_2)^3\,+\,...
\end{aligned}
\end{equation}
The trace in this expression is taken both over spinor and colour indices (for each gauge group). At this point it is useful to notice the following properties for gauge propagators:
\begin{equation}\label{prop identities}
\begin{aligned}
(\Delta_i)_\a^{\ \a}&=\frac{2(\kappa_iD^2+p^2)}{p^2(\kappa_i^2+p^2)}\equiv\delta_i,\\
(\Delta_1)_\a^{\ \b}(\Delta_1)_\b^{\ \g}&=\frac{2(\kappa_1D^2+p^2)}{p^2(\kappa_1^2+p^2)}(\Delta_1)_\a^{\ \g}=\delta_1(\Delta_1)_\a^{\ \g},\\
(\Delta_2)_\a^{\ \b}(\Delta_2)_\b^{\ \g}&=\frac{2(\kappa_2D^2+p^2)}{p^2(\kappa_2^2+p^2)}(\Delta_2)_\a^{\ \g}=\delta_2(\Delta_2)_\a^{\ \g},\\
(\Delta_1)_\alpha^{\ \beta}(\Delta_2)_{\b}^{\ \g}&=\delta_1(\Delta_2)_\a^{\ \g}=\delta_2(\Delta_1)_\a^{\ \g}.
\end{aligned}
\end{equation}
We can then repackage the expression in \eqref{sigma} in terms of the following matrices:
\begin{equation}
    \mathcal{M}\,=\,\left(
    \begin{matrix}
    M^{AB} && G^{AN}\\
    (G^\dagger)^{MB} && N^{MN}
    \end{matrix}
    \right),
\end{equation}
\begin{equation}
    \Delta\,=\,\left(
    \begin{matrix}
    \delta_1\cdot \mathbf{1}_{N_1\times N_1} && 0\\
    0 && \delta_2\cdot \mathbf{1}_{N_2\times N_2}
    \end{matrix}
    \right),
\end{equation}
where $N_1$ and $N_2$ are the dimensions of the fundamental representations of $G_1$ and $G_2$, respectively. At this point we can define the generic n-th 1-loop contribution as
\begin{equation}
\Sigma^{(n)}_{\text{1-loop}}\equiv\Tr(\mathcal M\Delta)^n
\end{equation}
and sum all the contributions to get
\begin{equation}
\begin{aligned}
    \Sigma&=\sum_n \mathcal S_n\ \Sigma^{(n)}_{\text{1-loop}}\\
    &=-\frac{1}{2}\Tr(\mathcal{M}\Delta)+\frac{1}{4}\Tr(\mathcal{M}\Delta\mathcal{M}\Delta)-\frac{1}{6}\Tr(\mathcal{M}\Delta\mathcal{M}\Delta\mathcal{M}\Delta)+\, ... \\
    &=-\frac{1}{2}\Tr\log(\mathbb {1}_{(N_1+N_2)\times(N_1+N_2)}+\mathcal M\Delta)\\
    &=-\frac{1}{2}\log\det(\mathbb {1}_{(N_1+N_2)\times(N_1+N_2)}+\mathcal M\Delta)\\
    &=-\frac{1}{2}\log\big(\det\left( (\mathbb{1}_{N_1\times N_1}+\delta_1 M)-\delta_1\delta_2 G(\mathbb{1}_{N_2\times N_2}+\delta_2 N)^{-1}G^T\right)\cdot\det\left(\mathbb{1}_{N_2\times N_2}+\delta_2 N \right)\big)
    \end{aligned}
    \end{equation}
where in the first step we used the fact that $\mathcal S_n= \frac{1}{n!}\left(-\frac{1}{2}\right)^n(2n-2)!!=\frac{(-1)^n}{2n}$.

We now switch to the case of $SU(2)\times SU(2)$ gauge group and note few properties intrinsic to this case, which are helpful to further simplify the result. In this special case $M$, $N$, and $G$ are all three by three diagonal matrices, therefore they all commute. In fact, the expressions in (\ref{matrices}) imply
\begin{equation}
\begin{aligned}
M^{AB}=\frac{g_1^2\Tr \bar{\Phi}\Phi}{4}\,\delta^{AB},\\
N^{AB}=\frac{g_2^2\Tr \bar{\Phi}\Phi}{4}\,\delta^{AB}.
\end{aligned}
\end{equation}
Hence we obtain
\begin{equation}\label{sigma1}
    \Sigma=-\frac{1}{2}\Tr\log \left[\mathbb {1}_{3\times3}+\delta_1M+\delta_2N+\delta_1\delta_2(MN-G^TG)\right].
\end{equation}
Taking into account that the eigenvalues of $G^T G$ are given by 
\begin{equation}
    \left(\frac{g_1^2g_2^2\det\Phi\det\bar{\Phi}}{4},\frac{g_1^2g_2^2\det\Phi\det\bar{\Phi}}{4},\frac{g_1^2g_2^2(\Tr\bar{\Phi}\Phi)^2}{16} \right),
\end{equation}
and introducing the notation
\begin{equation}
\rho\equiv\Tr\bar{\Phi}\Phi,\qquad B\equiv 2\det\Phi,\qquad \bar B\equiv 2\det\bar\Phi, 
\end{equation}
we can substitute the above results in \eqref{sigma1} and obtain
\begin{equation}
    \Sigma=-\log\left[1+\frac{\rho}{4}(g_1^2\delta_1+g_2^2\delta_2)+\frac{\rho^2-B\bar{B}}{16} g_1^2g_2^2\delta_1\delta_2 \right]-\frac{1}{2}\log\left[1+\frac{\rho}{4}(g_1^2\delta_1+g_2^2\delta_2)\right].
\end{equation}
At this point we can substitute everything in \eqref{superpot} and get the complete 1-loop expression for the superpotential:
\begin{equation}
    \begin{aligned}
    \mathcal{W}_{\text{1-loop}}=&-\int \frac{d^3 p}{(2\pi)^3}d^2\theta'\delta(\theta-\theta')\log\left[1+\frac{\rho}{4}(g_1^2\delta_1+g_2^2\delta_2)+\frac{\rho^2-B\bar{B}}{16} g_1^2g_2^2\delta_1\delta_2 \right]\delta^2(\theta'-\theta)\\
    &-\frac{1}{2}\int \frac{d^3 p}{(2\pi)^3}d^2\theta'\delta(\theta-\theta')\log\left[1+\frac{\rho}{4}(g_1^2\delta_1+g_2^2\delta_2) \right]\delta^2(\theta'-\theta).
    \end{aligned}
\end{equation}
Notice now that the expressions between the delta-functions are functions of $D^2$, which enters there via $\delta_{1,2}$. Using the last identity in (\ref{identities}), one can reduce an arbitrary function of $D^2$ to a linear function of $D^2$. Using then the identities
\begin{equation}
    \begin{aligned}
    &\delta^2(\theta-\theta')\delta^2(\theta'-\theta)=0,\\
    &\delta^2(\theta-\theta')D^{\alpha}\delta^2(\theta'-\theta)=0,\\
    &\delta^2(\theta-\theta')D^2\delta^2(\theta'-\theta)=\delta^2(\theta-\theta'),
    \end{aligned}
\end{equation}
we rewrite the effective superpotential as
\begin{equation}
    \begin{aligned}
    \mathcal{W}_{\text{1-loop}}=&-\int \frac{d^3 p}{(2\pi)^3}\log\left[1+\frac{\rho}{4}(g_1^2\delta_1+g_2^2\delta_2)+\frac{\rho^2-B\bar{B}}{16} g_1^2g_2^2\delta_1\delta_2 \right]\lvert_{D^2}\\
    &-\frac{1}{2}\int \frac{d^3 p}{(2\pi)^3}\log\left[1+\frac{\rho}{4}(g_1^2\delta_1+g_2^2\delta_2) \right]\lvert_{D^2}.
    \end{aligned}
\end{equation}
Here $\lvert_{D^2}$ means that we should reduce the functions of $D^2$ to linear ones, as explained above, and then take the coefficient in front of $D^2$. In fact, all the coefficients in front of $D^2$ terms can be easily extracted with the aid of the following identity
\begin{equation}
    (\kappa D^2+p^2)^n\lvert_{D^2}=\frac{1}{|p|}\text{Im}\left( (i\kappa|p|+p^2)^n \right),
\end{equation}
which leads us to the final result for the effective superpotential
\begin{equation}\label{effepotentialfinal}
    \begin{aligned}
    \mathcal{W}_{\text{1-loop}}=&-\int \frac{d^3 p}{(2\pi)^3}\frac{1}{|p|}\text{Im}\log\left[1+\frac{\rho}{2}\left(\frac{g_1^2}{(p^2-i\kappa_1|p|)}+\frac{g_2^2}{(p^2-i\kappa_2|p|)}\right)+\frac{g_1^2 g_2^2(\rho^2-B\bar{B})}{4(p^2-i\kappa_1|p|)(p^2-i\kappa_2|p|)} \right]\\
    &-\frac{1}{2}\int \frac{d^3 p}{(2\pi)^3}\frac{1}{|p|}\text{Im}\log\left[1+\frac{\rho}{2}\left(\frac{g_1^2}{(p^2-i\kappa_1|p|)}+\frac{g_2^2}{(p^2-i\kappa_2|p|)}\right) \right].
    \end{aligned}
\end{equation}
In the study of SUSY vacua, the derivatives of $\mathcal{W}_{\text{eff}}$ with respect to $\rho$, $B$ and $\bar{B}$ are going to be relevant. More precisely, we will need their asymptotic behaviour at large field values. For the general case of $k_1\neq-k_2 $ they are computed to be
\begin{equation}\label{asymptotics1}
\begin{aligned}
    \partial_{\rho}\mathcal{W}_{\text{1-loop}}\,&=
    \begin{cases}
       -\frac{F_1+F_4}{\rho^{1/2}} \,+\, \mathcal O \left( \rho^{-3/2}, (\frac{B}{\rho})^2 \right)\quad \text{if}\ \frac{B}{\rho}\ll1, \\
      -G-\frac{3F_1+F_2}{\rho^{1/2}}\,+\,\mathcal O \left( \rho^{-3/2} \right)\ \quad \text{if}\ B=\rho
    \end{cases}\\
    \partial_{B}\mathcal{W}_{\text{1-loop}}\,&=
    \begin{cases}
      \frac{F_3 B}{\rho^{3/2}} \,+\,B\, \mathcal O \left( \rho^{-5/2}, \frac{B}{\rho} \right)\ \quad \text{if}\ \frac{B}{\rho}\ll1, \\
      G + \frac{F_2}{\rho^{1/2}}+\, \mathcal O \left( \rho^{-3/2} \right) \qquad\text{if}\ B=\rho.
    \end{cases}
\end{aligned}
\end{equation}
Here $F_1, F_2, F_3, F_4, G_1$ are functions of $g_1,g_2,\kappa_1,\kappa_2$:
\begin{subequations}\label{asymptotics2}
\begin{align}
F_1&=\frac{\kappa_1g_1^2+\kappa_2g_2^2}{16\sqrt{2}\pi\sqrt{g_1^2+g_2^2}},\\
F_2&=\frac{g_1^2g_2^2\left((\kappa_1+\kappa_2)(g_1^2+g_2^2) - 3(\kappa_1 g_2^2+\kappa_2 g_1^2)\right)}{4\sqrt{2}\pi(g_1^2+g_2^2)^{5/2}},\\
F_3&=\frac{g_1g_2(\kappa_1g_2+\kappa_2g_1)}{4\sqrt{2}\pi(g_1+g_2)^2},\\
F_4&=\frac{g_1 \kappa_1+g_2 \kappa_2}{8 \sqrt{2} \pi }\\
G&=\frac{g_1^2g_2^2}{4\pi(g_1^2+g_2^2)}.
\end{align}
\end{subequations}

For the case of $k_1=-k_2$ one can go a bit further and get not only the leading asymptotics, but complete closed form expressions for $\partial_{\rho}\mathcal{W}_{\text{1-loop}}$ and $\partial_{|B|}\mathcal{W}_{\text{1-loop}}$ in the limits $B\rightarrow\rho$ and  $B\rightarrow 0$: these results are collected in Appendix \ref{Crocodiles}.

As is explained in \cite{Bashmakov:2018wts}, the asymptotic behaviour of the effective superpotential is fully determined by the one-loop contributions and does not receive higher-order corrections. This fact implies that the results in \eqref{asymptotics1},\eqref{asymptotics2} are actually exact.


\section{Phase diagrams of $SU(2)\times SU(2)$ models}\label{SUSU}

We are now ready to apply the results of the previous section to the study of the IR phases of the $SU(2)\times SU(2)$ quiver theories. We will add the tree-level mass term to the superpotential, such that the total superpotential is given by the sum of the mass term and the one-loop effective superpotential,
\begin{equation}
    \mathcal{W}=m\Tr \bar{\Phi}\Phi+\mathcal{W}_{1-loop}.
\end{equation}
With the effective superpotential at hands, we can proceed and study supersymmetric vacua of the theory, which are given by the critical points of the superpotential,
\begin{equation}\label{susyvacua}
    \bar{\partial}\mathcal{W}\,=\,0.
\end{equation}
As was discussed above \eqref{ClassicalModuliSpaceSU2}, we can apply two gauge transformations and put the scalar matrix into the form
\begin{equation}
    \Phi=\left(
    \begin{matrix}
    \phi_{11} && 0\\
    0 && \phi_{22}
    \end{matrix}
    \right),
\end{equation}
where the common phase of $\phi_{11}$ and $\phi_{22}$ have been set to zero thanks to the $U(1)$ baryonic symmetry. The equation in \eqref{susyvacua} can be expanded as
\begin{equation}
\begin{aligned}
    \phi_{11}\partial_{\rho}\mathcal{W}\,+\,\phi_{22}\partial_{|B|}\mathcal{W}\,=\,0,\\
    \phi_{22}\partial_{\rho}\mathcal{W}\,+\,\phi_{11}\partial_{|B|}\mathcal{W}\,=\,0.
\end{aligned}
\end{equation}
From the equations above one can then easily infer the following possibilities:
\begin{itemize}
    \item[1)] The vacuum at the origin, $\phi_{11}=\phi_{22}=0$.
    \item[2)] $\partial_{\rho} \mathcal{W}=-\partial_{|B|} \mathcal{W}\neq 0$, $\phi_{11}=\phi_{22}$. \footnote{There is also a possibility that $\partial_{\rho} \mathcal{W}=\partial_{|B|} \mathcal{W}\neq 0$, $\phi_{11}=-\phi_{22}$, but this is gauge-equivalent to the one stated above.}
    \item[3)] $\partial_{\rho}\mathcal{W}=\partial_{|B|}\mathcal{W}=0$.
\end{itemize}

At this point it is convenient to split the discussion into two different cases: Chern-Simons levels of the same sign and Chern-Simons levels of opposite signs.

\subsection{Chern-Simons levels of the same sign}\label{SameSignCSlevels}
In this subsection we deal with $\mathcal{N}=1$ $SU(2)_{k_1}\times SU(2)_{k_2}$ coupled to a bi-fundamental multiplet, and assume that $0\leq k_1\leq k_2$\footnote{The situation of $0\leq k_2\leq k_1$ is obtained by exchanging two nodes and the situation of $0\geq k_2\geq k_1$ is obtained by applying the time reversal transformation.}. We postpone the study of the $k_1=k_2=0$ case, which requires a separate treatment. When the mass parameter is large and positive, we can integrate the matter out and obtain a single vacuum with the infrared theory given by
\begin{equation}
    \mathcal{N}=1\qquad SU(2)_{k_1+1}\times SU(2)_{k_2+1}.
\end{equation}
This theory preserves supersymmetry and further flows to a topological CS theory in the IR (see Section \ref{ReviewVectorMultiplet}),
\begin{equation}
    SU(2)_{k_1}\times SU(2)_{k_2}\quad \text{TQFT}.
\end{equation}
The Witten index is (taking into account that there are six negative-mass Majorana gaugini)
\begin{equation}
    \text{WI}_+=(k_1+1)(k_2+1).
\end{equation}
If the mass parameter is large and negative, we can again integrate the matter out, but with the result
\begin{equation}
    \mathcal{N}=1\qquad SU(2)_{k_1-1}\times SU(2)_{k_2-1}.
\end{equation}
The fate of this vacuum now depends on the values of $k_1$ and $k_2$.
\begin{itemize}
    \item If $k_1,k_2>1$, the vacuum preserves supersymmetry and flows to a CS theory.
    \begin{equation}\label{theory1}
        SU(2)_{k_1-2}\times SU(2)_{k_2-2} \quad \text{TQFT},
    \end{equation}
    \begin{equation}
    \text{WI}_-=(k_1-1)(k_2-1).
\end{equation}
    \item If $k_1=1$, $k_2>1$, the vacuum breaks supersymmetry, and we get a Majorana goldstino together with a decoupled CS theory,
    \begin{equation}
        G_{\alpha}\,+\,U(1)_2\times SU(2)_{k_2-2} \quad\text{TQFT}.
    \end{equation}
    \item If $k_1=k_2=1$, the vacuum breaks supersymmetry, and the IR theory is
    \begin{equation}
        G_{\alpha}\,+\,U(1)_2\times U(1)_2\quad\text{TQFT}.
    \end{equation}
    \item If $k_1=0$, $k_2>1$, supersymmetry is preserved, and the IR theory is
    \begin{equation}
        SU(2)_{k_2-2}\quad\text{TQFT},
    \end{equation}
    \begin{equation}
        \text{WI}_-=-(k_2-1).
    \end{equation}
    \item If $k_1=0$, $k_2=1$, supersymmetry is again broken, and we get in the IR
    \begin{equation}
        G_{\alpha}\,+\,U(1)_2\quad\text{TQFT}.
    \end{equation}
\end{itemize}
We note in particular that the large positive mass phase and the large negative mass phase have different Witten indices. In order to understand the transition between the large negative mass phase and the large positive mass phase, it is useful to understand the dynamics near the point $m=0$, where the asymptotic behaviour of the superpotential changes and the Witten index can jump.

We start by observing that the vacuum of the first kind, namely at the origin of the field space, $\phi_{11}=\phi_{22}=v_1=0$, exists for $m=0$ as well as for $m$ small and positive or small and negative. We identify this vacuum with the semiclassical vacuum we have seen at large and negative mass. (It will be evident in a moment that this identification leads to a consistent phase diagram providing the matching of the Witten index at the phase transition locus). This vacuum either preserves supersymmetry or, if one of the CS levels is equal to one, breaks it non-perturbatively. It is then expected to find new vacua appearing from the infinity of the field space near the line $m=0$, and whose total Witten index must be different from zero. We thus initiate the search of these vacua, which must be either of the second or of the third type.

\subsubsection{Non-Abelian vacuum}
We first turn to the analysis of the vacuum of the second kind, with $\phi_{11}=\phi_{22}=v_2$ for some real and positive $v_2$, and with $\rho=B=\bar{B}=2v_2^2$. Then the equation $\partial_{\rho}\mathcal{W}=-\partial_{|B|}\mathcal{W}$ turns into
\begin{equation}\label{eqvev1}
-G-\frac{3F_1+F_2}{\rho^{1/2}}+m=-G -\frac{F_2}{\rho^{1/2}}
\end{equation}
We immediately notice that, thanks to the positivity of $F_1$ (see \eqref{asymptotics2}), we must necessarily have $m>0$. Therefore, when we move from the negative-mass region and cross the $m=0$ line, a new vacuum is found, in precise accordance with our expectations. The expression for the vev can be then easily found and is given by
\begin{equation}\label{SUSUfirstvacuum}
 v_2= \frac{3F_1}{\sqrt 2 m}
\end{equation}
Next we determine the effective low-energy theory of this vacuum. The vev
\begin{equation}
\Phi=\left(
\begin{matrix}
v_2 && 0\\
0 && v_2
\end{matrix}
\right)
\end{equation}
breaks the global baryonic symmetry $U(1)_B$, thus we expect to see the corresponding Goldstone boson with its superpartner. The vacuum also breaks the gauge group to the diagonal $SU(2)$ with the induced CS level equal to $k_1+k_2$. The CS level can receive quantum corrections when massive fermions charged under the unbroken $SU(2)$ are integrated out, so we need to understand the fermionic mass spectrum.

The fermionic mass terms originate from the superpotential, from the gaugini-matter coupling terms, and from the gaugini mass term, a supersymmetric counterpart of the CS term:
\begin{equation}\label{fermionmasses}
\begin{aligned}
   \mathcal L_{\psi^2}=& \frac{\partial^2 \mathcal{W}}{\partial{\bar{\Phi}_{\hat i j}\partial\Phi_{k\hat l}}}\bar{\Psi}_{\hat i j}\Psi_{k\hat l}\,+\frac{1}{2}\left(\,\frac{\partial^2 \mathcal{W}}{\partial{\bar{\Phi}_{\hat ij}\partial\bar{\Phi}_{\hat kl}}}\bar{\Psi}_{\hat ij}\Psi^c_{\hat k l}\,+\,\text{c.c.}\right)\,+\,\left(i g_1 \text{Tr}\bar{\Psi}\lambda_1 \Phi\,+\,i g_2\text{Tr}\bar{\Psi}^c\lambda_2 \bar{\Phi}\,+\,\text{c.c.}\right)+\\
   &-\kappa_1\text{Tr}\bar{\lambda}_1\lambda_1\,-\,\kappa_2\text{Tr}\bar{\lambda}_2\lambda_2,
\end{aligned}
\end{equation}
where the indices are put on the same line for convenience, and scalars are assumed to take their vev.

While we are given fermions in representations of $SU(2)\times SU(2)$ group, it is natural to decompose them into representations of the diagonal $SU(2)$ group. The decomposition goes as follows,
\begin{equation}\label{SecondVacuumFermionsDecomposition}
    \Psi_{i}^{\ \hat j}=\frac{1}{2}\left[\psi^a_{\text{Re}}(\sigma^a)_{i}^{\ \hat j}+i\psi^a_{\text{Im}}(\sigma^a)_{i}^{\ \hat j}+(\psi_0+i\psi_G)\delta_i^{\ \hat j}\right],
\end{equation}
where we introduced two Majorana modes, $\psi_0$ and $\psi_G$, neutral under the diagonal $SU(2)$, and two Majorana multiplets, $\psi_{\text{Re}}^a$ and $\psi_{\text{Im}}^a$, transforming in the adjoint representation of $SU(2)$. Two other adjoint multiplets are provided by $\lambda_1$ and $\lambda_2$.\\
It follows from \eqref{fermionmasses} that the mass of $\psi_0$ and $\psi_G$ are
\begin{subequations}
\begin{align}
m_0&=4v_2^2\left(\partial^2_\rho\mathcal{W}+2\partial_\rho\partial_B\mathcal{W}+\partial^2_B\mathcal{W}\right)=m,\label{m0}\\
m_G&=0,\label{mg}
\end{align}    
\end{subequations}
thus we can identify the latter with the superpartner of the Goldstone boson associated with the broken $U(1)_B$. The mass of $\psi_{\text{Re}}$ instead reads
\begin{equation}
    m_{\text{Re}}=2\partial_\rho\mathcal{W}=-2G-2m\left(1+\frac{F_2}{3F_1}\right).
\end{equation}
Notice that, since we must have $m>0$ from \eqref{eqvev1} and $G>0$ by definition, it follows that $m_{\text{Re}}$ must be negative.\\ Finally, $\psi_{\text{Im}}$, $\lambda_1$, and $\lambda_2$ mix with each other via the mass matrix
\begin{equation}\label{SecondVacuumFermionsMixing}
   \left(
    \begin{matrix}
    0 && g_1 v_2 && -g_2 v_2\\
    g_1 v_2 && -\kappa_1 && 0\\
    -g_2 v_2 && 0 && -\kappa_2 
    \end{matrix}
    \right).
\end{equation}
This mass matrix has one positive eigenvalue and two negative eigenvalues (one of the eigenmodes with a negative eigenvalue can be identified with the gaugino of the unbroken gauge group). Therefore, in total we have three adjoint multiplets with negative mass and one adjoint multiplet with positive mass, coupled to the unbroken $SU(2)$: integrating them out, we get a shift of the CS level equal to $-2$.

We are thus ready to formulate the infrared dynamics of this vacuum. 
Unless $k_1=0$ and $k_2=1$, it preserves supersymmetry, and the infrared dynamics is described by
\begin{equation}
    \Phi_G\, +\, SU(2)_{k_1+k_2-2}\quad\text{TQFT}, 
\end{equation}
where $\Phi_G$ is the Goldstone supermultiplet.
If, on the contrary, $k_1=0$ and $k_2=1$, supersymmetry is spontaneously broken, and we get in the infrared
\begin{equation}\label{NewNon-SusyVacuum}
    \phi_G\,+\,G_{\alpha}\,+\,U(1)_2\quad\text{TQFT},
\end{equation}
with $\phi_G$ being the Goldstone boson.

\subsubsection{Abelian vacuum}\label{SameCSLevelsThirdVac}
Next we consider the vacuum of the third kind appearing from infinity in the field space. The condition $\partial_B\mathcal{W}=0$ requires that $B=0$\footnote{One can first observe that $B\leq\rho$, and from \eqref{asymptotics1} we see that for $B=\rho\rightarrow\infty$, $\partial_B\mathcal{W}>0$, while for $B=0$, $\partial_B\mathcal{W}=0$, as desired. It can then be examined numerically that there are no other solutions for $0<B<\rho$.}. The condition $\partial_{\rho}\mathcal{W}=0$ then reads as
\begin{equation}
    m-\frac{F_1+F_4}{\rho^{1/2}}=0.
\end{equation}
Since $F_1+F_4>0$, there are no solutions for $m\leq0$, while for $m>0$ we find 
\begin{equation}
    \Phi=\left(
    \begin{matrix}
    v_3 && 0\\
    0 && 0
    \end{matrix}
    \right),\qquad v_3=\frac{F_1+F_4}{m}.
\end{equation}

The next step is to study the IR dynamics of this vacuum. The $U(1)_B$ is preserved here, and is generated by \footnote{We note that this choice of the preserved $U(1)_B$ is not unique, and is defined up to an action of the gauge transformation.}
\begin{equation}
    \frac{i}{2}\left(
    \begin{matrix}
    1 && 0 \\
    0 && 1
    \end{matrix}
    \right)_1
    \oplus
    \frac{i}{2}\left(
    \begin{matrix}
    1 && 0 \\
    0 && -1
    \end{matrix}
    \right)_2,
\end{equation}
and the gauge symmetry is broken to $U(1)$, generated by
\begin{equation}
\frac{i}{2}
\left(
    \begin{matrix}
    1 && 0 \\
    0 && -1
    \end{matrix}
    \right)_1
    \oplus
    \frac{i}{2}
    \left(
    \begin{matrix}
    1 && 0 \\
    0 && -1
    \end{matrix}
    \right)_2.
\end{equation}
where the subscripts $1,2$ indicate respectively the first or the second gauge group factor. This Abelian gauge field inherits the Chern-Simons level $2(k_1+k_2)$.\\
We can now classify fermionic modes according to their charges with respect to the unbroken $U(1)\times U(1)_B$. The matter multiplet fermions and two types of gaugini can be decomposed as
\begin{equation}
    \Psi=\left(
    \begin{matrix}
    \frac{\eta+i\chi}{\sqrt 2} && \psi_+\\
    \psi_- && \psi_0
    \end{matrix}
    \right),\qquad
    \lambda_1=\frac{1}{2}\left(
    \begin{matrix}
    \lambda_{1,0} && \sqrt{2}\lambda_{1,+}\\
    \sqrt{2}\lambda^c_{1,+} && -\lambda_{1,0}
    \end{matrix}
    \right),\qquad
    \lambda_2=\frac{1}{2}\left(
    \begin{matrix}
    \lambda_{2,0} && \sqrt{2}\lambda_{2,+}\\
    \sqrt{2}\lambda^c_{2,+} && -\lambda_{2,0}
    \end{matrix}
    \right).
\end{equation}
All the types of the modes (Majorana or Dirac) as well as their charges are summarized in Table \ref{tab:my_label}. 
\begin{table}[]
    \centering
    \renewcommand\arraystretch{1.2} 
    \begin{tabular}{||c|c|c|c|c|c|c|c|c|c||}
    \hline 
    Mode &  $\eta$ & $\chi$ & $\psi_0$ & $\psi_+$ & $\psi_-$ & $\lambda_{1,0}$ & $\lambda_{1,+}$ &$\lambda_{2,0}$ & $\lambda_{2,+}$\\
    \hline
    Type &  \textbf M &  \textbf M & \textbf D & \textbf D & \textbf D &  \textbf M &  \textbf D & \textbf M &  \textbf D  \\
    $U(1)$ &  0 & 0 & 0 & 1 & -1 & 0 & 1 & 0 & 1\\
    $U(1)_B$ & 0 & 0 & 1 & 1 & 0 & 0 & 0 & 0 & 1\\
    \hline
    \end{tabular}
    \caption{Fermion modes}
    \label{tab:my_label}
\end{table}
The masses of fermions neutral under the $U(1)$ gauge group are determined as follows. By starting again from \eqref{fermionmasses}, we get that $\eta$ and $\psi_0$ masses come from the superpotential,
\begin{equation}\label{ThirdVacuumNeutralDiagonal}
    \begin{aligned}
    &m_{\eta}=2v_3^2\frac{\partial^2\mathcal{W}}{\partial\rho^2}=m,\\
    &m_{\psi_0}=v_3^2\left(\frac{\partial^2\mathcal{W}}{\partial B^2}+\frac{1}{B}\frac{\partial\mathcal{W}}{\partial B}\right)=2v_3^2\frac{\partial^2\mathcal{W}}{\partial B^2}=2m\frac{F_3}{F_1+F_4},
    \end{aligned}
\end{equation}
whereas $\chi$, $\lambda_{1,0}$, $\lambda_{2,0}$ mix with each other through the mass matrix
\begin{equation}\label{ThirdVacuumNeutralMixing}
    \left(
    \begin{matrix}
    0 && \frac{1}{\sqrt 2}g_1 v_3 && -\frac{1}{\sqrt 2}g_2 v_3\\
    \frac{1}{\sqrt 2}g_1 v_3 && -\kappa_1 && 0\\
    -\frac{1}{\sqrt 2}g_2 v_3 && 0 && -\kappa_2 
    \end{matrix}
    \right).
\end{equation}
There are also mixing modes charged under the $U(1)$ gauge group, in particular, $\psi_{-}$ mixes with $\lambda^c_{1,+}$ via
\begin{equation}\label{ThirdVacuumChargedMixingOne}
   \left(
    \begin{matrix}
    0 && i  \frac{1}{\sqrt 2}g_1 v_3 \\
    -i \frac{1}{\sqrt 2}g_1 v_3 && -\kappa_1
    \end{matrix}
    \right),
\end{equation}
and $\psi_{+}$ mixes with $\lambda_{2,+}$ via
\begin{equation}\label{ThirdVacuumChargedMixingTwo}
    \left(
    \begin{matrix}
    0 && -i\frac{1}{\sqrt 2} g_2 v_3 \\
    i \frac{1}{\sqrt 2}g_2 v_3 && -\kappa_2
    \end{matrix}
    \right).
\end{equation}
Both matrices have one positive eigenvalue and one negative eigenvalue, which implies that the $U(1)$ CS level does not get renormalized when these massive modes are integrated out. We thus conclude that at low energies we get a pure Abelian CS theory
\begin{equation}
    U(1)_{2(k_1+k_2)}\quad \text{TQFT}.
\end{equation}

\vspace{10pt}
\begin{center}
$ {\ast}\,{\ast}\,{\ast} $
\end{center}

Few comments are in order. First, we were able to follow the appearance of two new vacua as far as the line $m=0$ is crossed. This process is controlled just by the leading asymptotic of the effective superpotential, which in turn is determined only by the one-loop contribution. We thus conclude that we have rigorously derived the existence of these vacua. Second, there might be supersymmetric vacua emerging for some values of the parameters $g_1,g_2,m$, which don't come from infinity, but rather appear at finite field values. These vacua should have vanishing total Witten index, and their dynamics is a priori governed by perturbation theory at all orders, and not just at one-loop level. We do not have reliable tools to study them and, moreover, there are no consistency requirements (i.e. Witten index matching) that would necessitate their existence.

\vskip 30pt

\subsubsection{Summary of the results and Phase Diagrams}
\begin{figure}[h]
    \centering
    \includegraphics[width=0.45\textwidth]{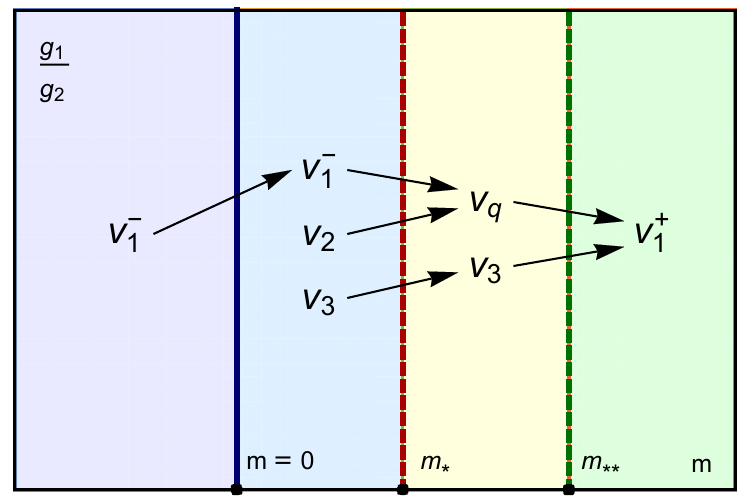}
    \caption{Structure of the phase diagrams of the $SU(2)_{k_1}\times SU(2)_{k_2}$ quivers and, as will be discussed later, the $SU(2)_{k_1}\times U(2)_{k_2, k_3}$ quivers with either $k_1,k_2>1$, or $k_1=0$, $k_2>1$. Dashed lines correspond to the second order phase transitions, while the solid line is the wall. Supersymmetric vacua at each phase are indicated.}
    \label{PhaseDiag1}
\end{figure}

We are now able to formulate the phase diagram of the theory under consideration.  We start with a generic case of $k_1>1$ and $k_2>1$, and the relevant phase diagram is schematically depicted in Figure \ref{PhaseDiag1}, where we attempt to reflect only the topology of the phase diagram\footnote{In particular, various straight lines appearing on the figure should in practice be curved.}. At large and negative masses, and up to the $m=0$ line there is the supersymmetric semiclassical vacuum described by
\begin{equation}
    SU(2)_{k_1-2}\times SU(2)_{k_2-2}\quad\text{TQFT},
\end{equation}
as we saw in \eqref{theory1}. This vacuum is denoted on the figure by $v_1^-$, and this phase corresponds to the purple region. The Witten index of this vacuum is
\begin{equation}
    \text{WI}_1=(k_1-1)(k_2-1).
\end{equation}
When we cross the wall at $m=0$, two new vacua come in from infinity,
\begin{equation}
    \begin{aligned}
    &\Phi_G\, +\, SU(2)_{k_1+k_2-2}\quad\text{TQFT},\quad\text{WI}_2=0,\\
    &U(1)_{2(k_1+k_2)}\quad \text{TQFT},\quad\text{WI}_3=2(k_1+k_2),
    \end{aligned}
\end{equation}
 giving us a phase with three vacua (the new vacua are $v_2$ and $v_3$, and the corresponding region in Figure \ref{PhaseDiag1} is the light blue one). These three vacua must undergo, generically, two second-order phase transitions (lines $m_*$ and $m_{**}$, which are actually functions of $g_1,g_2$), merging into a single vacuum seen at large positive masses which we indicate with $v_1^+$. 
 \begin{wrapfigure}{r}{0.22\textwidth}
 \includegraphics[width=0.30\textwidth]{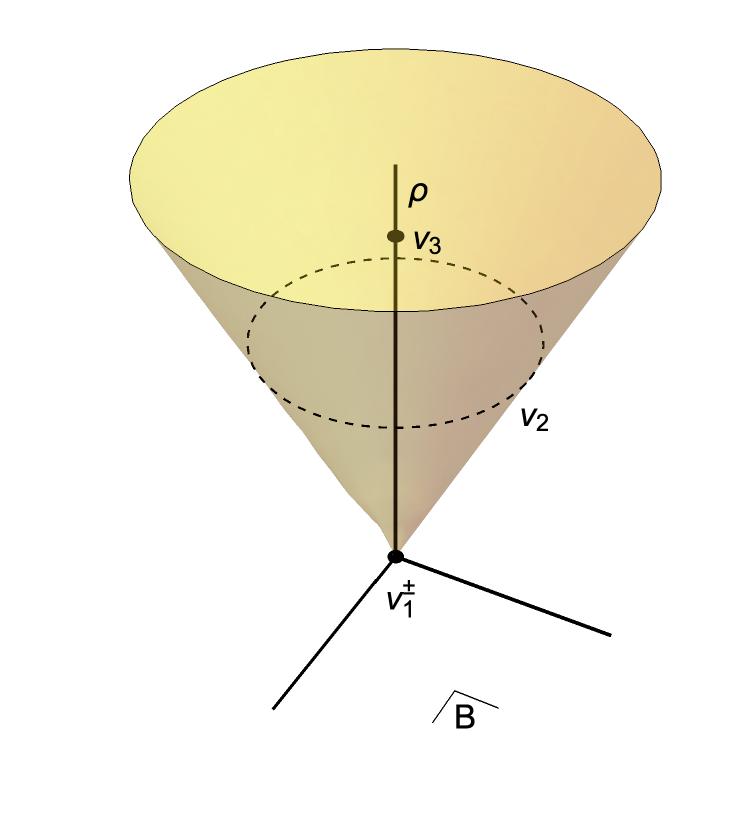}
 \end{wrapfigure}
 These phase transitions are supposed to happen somewhere around the origin of the field space, where the physics is strongly coupled, and we do not have much control over it. The structure of vevs in each vacuum suggests (see the figure on the right) that at the first phase transition either $v_2$ merges with $v_1^-$, or $v_3$ merges with $v_1^-$: we conjecture, basing on a duality proposed in Section \ref{Dualities}, that the first option is realized. In the intermediate phase we then still get the Abelian vacuum $v_3$ and some other vacuum, $v_q$, which is guessed to support again the $SU(2)_{k_1-2}\times SU(2)_{k_2-2}$ CS theory (the yellow region in Figure \ref{PhaseDiag1}). At the second phase transition two vacua merge and produce the large positive mass vacuum $v_1^+$.  

Let us now discuss some exceptional cases with low values of the CS levels.
\begin{itemize}
    \item If $k_1=1$, $k_2\geq1$, the negative mass vacuum breaks supersymmetry. As soon as the wall $m=0$ is crossed, two supersymmetric vacua appear,
\begin{equation}
    \begin{aligned}
    &\Phi_G\, +\, SU(2)_{k_2-1}\quad\text{TQFT},\quad\text{WI}_2=0,\\
    &U(1)_{2(k_2+1)}\quad \text{TQFT},\quad\text{WI}_3=2(k_2+1),
    \end{aligned}
\end{equation}
at some value of the mass parameter they merge and give rise to the large mass vacuum,
\begin{equation}
  SU(2)_{1}\times SU(2)_{k_2}\quad\text{TQFT}.
\end{equation}
The picture is illustrated in Figure \ref{PhaseDiag2}.

\begin{figure}[h]
    \centering
    \includegraphics[width=0.45\textwidth]{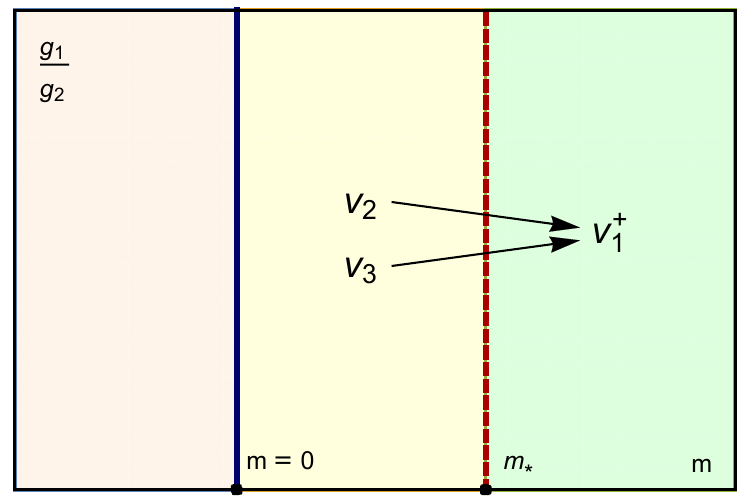}
    \caption{Phase diagram for $k_1=1$, $k_2\geq1$. Dashed line correspond to the second order phase transition, while the solid line is the wall. Supersymmetric vacua at each phase are indicated.}
    \label{PhaseDiag2}
\end{figure}

    \item If $k_1=0$ and $k_2>1$, the negative mass vacuum is supersymmetric and the theory flows to
    \begin{equation}
        SU(2)_{k_2-2}\quad\text{TQFT},\quad\text{WI}_0=-(k_2-1).
    \end{equation}
    When we cross the line $m=0$, two supersymmetric vacua appear:
    \begin{equation}
    \begin{aligned}
    &\Phi_G\, +\, SU(2)_{k_2-2}\quad\text{TQFT},\quad\text{WI}_2=0,\\
    &U(1)_{2k_2}\quad \text{TQFT},\quad\text{WI}_3=2k_2,
    \end{aligned}
    \end{equation}
    When $m$ is increased, the resulting three vacua undergo two phase transitions and produce a supersymmetric vacuum with
    \begin{equation}
        SU(2)_{k_2}\quad\text{TQFT}.
    \end{equation}
    
    \begin{figure}[h]
    \centering
    \includegraphics[width=0.45\textwidth]{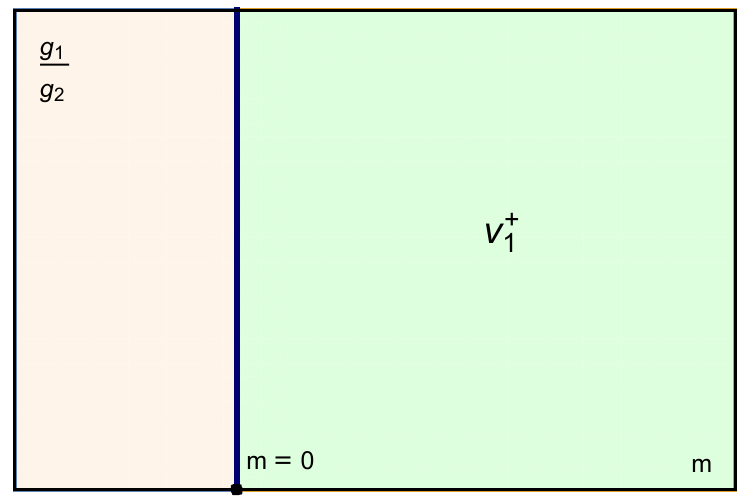}
    \caption{Phase diagram for $k_1=0$, $k_2=1$. The solid line correspond to the wall. The supersymmetric vacuum is indicated.}
    \label{PhaseDiag3}
\end{figure}

    \item If $k_1=0$ and $k_2=1$, the negative mass vacuum is not supersymmetric. When the wall is crossed, there appears one SUSY-breaking vacuum (see \eqref{NewNon-SusyVacuum}) and one supersymmetric vacuum with
    \begin{equation}
        U(1)_2\quad\text{TQFT}
    \end{equation}
in the IR. Using the chain of dualities for TQFTs, we observe
\begin{equation}\label{LRduality2}
    U(1)_2\ \longleftrightarrow\  U(1)_{-2}\longleftrightarrow\ SU(2)_1,
\end{equation}
and so the new supersymmetric vacuum came in from infinity exactly reproduces the semiclassical large mass vacuum. Therefore, in this case the theory does not undergo any phase transitions. (See Figure \ref{PhaseDiag3})

\end{itemize}

An important check of the picture that we are suggesting here is the matching of the Witten indices upon the phase transition points. As an example, in the case of $k_1,k_2>1$ we have
\begin{equation}
    \text{WI}_1+\text{WI}_2+\text{WI}_3=(k_1-1)(k_2-1)+0+2(k_1+k_2)=(k_1+1)(k_2+1).
\end{equation}
The right-hand side is nothing else but the Witten index of the large mass vacuum.

\subsection{Chern-Simons levels of opposite signs}\label{SUSUCSlevelsofOppositeSigns}

Having understood the phase diagram for the case when two CS levels have the same sign, we now turn to the situation when the first CS level is positive and the second is negative. We thus consider the theory
\begin{equation}
    SU(2)_{k_1}\times SU(2)_{-k_2},
\end{equation}
coupled as before to a bi-fundamental matter $\Phi$, with $k_1,k_2>0$, and we will also assume w.l.g. that $k_1>k_2$. The results of the following discussion are summarized in Figure \ref{PhaseDiag4}.

As before, it is useful to start the analysis by considering the large mass semiclassical phases. For large and negative masses we find a supersymmetric vacuum with
\begin{equation}
    SU(2)_{k_1-2}\times SU(2)_{-k_2}\quad\text{TQFT}
\end{equation}
in the IR ($v_1^-$ in Figure \ref{PhaseDiag4}). In the large positive mass phase we see the following picture:
\begin{itemize}
    \item When $k_2>1$, we get a supersymmetric vacuum ($v_1^+$ on Figure \eqref{PhaseDiag4}) hosting a CS theory,
    \begin{equation}
        SU(2)_{k_1}\times SU(2)_{-k_2+2}\quad\text{TQFT}.
    \end{equation}
    \item When $k_2=1$, SUSY gets broken, and the IR theory is given by
    \begin{equation}
        G_{\alpha}+ SU(2)_{k_1}\times U(1)_{2} \quad\text{TQFT}.
    \end{equation}
\end{itemize}
We first discuss in details the case of $k_2>1$, and then comment on the changes in the picture when $k_2=1$. 

As soon as the large mass phases are understood, the next step is to study the behaviour near the wall, at $m=0$. Again, we see that the $\phi_{11}=\phi_{22}=0$ vacuum exists on both sides of the wall. It will again be natural to identify this vacuum with one of the large mass vacua, however this time it is less obvious to decide which of the two should be chosen. We also remark that, while moving along the line $m=0$, three special points can be distinguished. These are the points for which the asymptotic behaviour of the effective superpotential \eqref{asymptotics1} changes in a certain way, and are given by 
\begin{subequations}
\begin{align}
    &F_1=0,\quad\frac{g_1}{g_2}=\left(\frac{k_2}{k_1}\right)^{\frac{1}{4}},\\
    &F_1+F_4=0,\quad\frac{g_1}{g_2}=\alpha,\\
    &F_3=0,\quad\frac{g_1}{g_2}=\frac{k_2}{k_1}.
\end{align}
\end{subequations}
Here $\alpha$ is the single positive root of the equation $x^4-\frac{k_2}{k_1}+2(x^3-\frac{k_2}{k_1})\sqrt{x^2+1}=0$, and we note that for $k_1>k_2$ we have
\begin{equation}\label{alpharelations}
\frac{k_2}{k_1}<\alpha<\left(\frac{k_2}{k_1}\right)^{\tfrac{1}{4}}
\end{equation}
Now, following the route of the previous subsection, we give a detailed discussion of the vacuum structure near the wall.

\subsubsection{Non-Abelian vacuum}

\begin{figure}[h]
    \centering
    \includegraphics[width=0.6\textwidth]{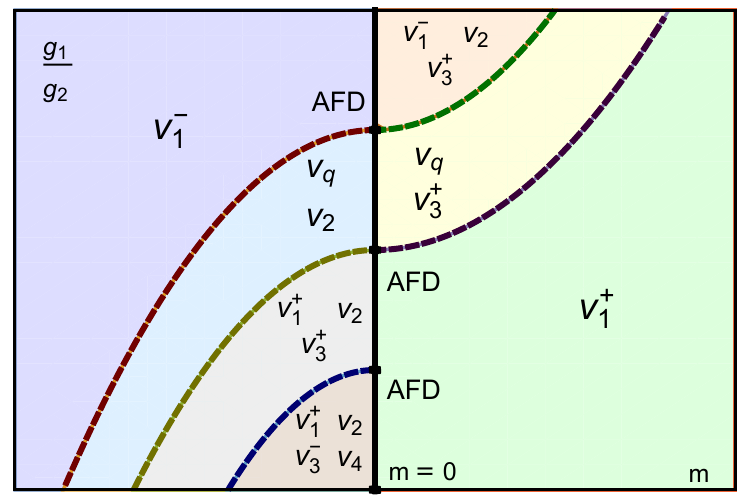}
    \caption{Structure of the phase diagrams of the $SU(2)_{k_1}\times SU(2)_{-k_2}$ quivers and, as will be clear later, the $SU(2)_{k_1}\times U(2)_{-k_2,-k_3}$ quivers with $k_1,k_2>0$ and $k_1>k_2$. Dashed lines correspond to the second order phase transitions, while the solid line is the wall. Supersymmetric vacua at each phase are indicated.}
    \label{PhaseDiag4}
\end{figure}

We first search for the vacuum of the second type. The equation $\rho=B=\bar{B}=2v_2^2$ again reduces to
\begin{equation}
    -\frac{3F_1}{\rho^{1/2}}+m=0,
\end{equation}
but now $F_1$ changes the sign when $g_1, g_2$ are varied. Indeed, what happens is that
\begin{equation}
    F_1\gtrless0,\quad\frac{g_1}{g_2}\gtrless\left(\frac{k_2}{k_2}\right)^{1/4}.
\end{equation}
We thus conclude that the vacuum of the second type is still given by
\begin{equation}
    v_2=\frac{3F_1}{\sqrt{2}m},
\end{equation}
and exists when either $m>0,\,\tfrac{g_1}{g_2}>\left(\tfrac{k_2}{k_2}\right)^{1/4}$ or $m<0,\,\tfrac{g_1}{g_2}<\left(\tfrac{k_2}{k_2}\right)^{1/4}$ ($v_2$ in the orange, light blue, grey and brown regions of Figure \ref{PhaseDiag4}). At the point $\tfrac{g_1}{g_2}=\left(\tfrac{k_2}{k_2}\right)^{1/4}$ the quantum potential develops an asymptotic direction with zero energy. This is the first special point mentioned above.

The gauge and global symmetry breaking pattern in this vacuum is the same as for the CS levels of the same sign, $U(1)_B$ is spontaneously broken, and the unbroken gauge group is $SU(2)_{k_1-k_2}$. We can also apply the previously obtained results for the fermionic mass spectrum, which does not undergo any changes. The resulting low-energy theory again depends on the values of the levels.
\begin{itemize}
    \item When $k_1>k_2+1$, supersymmetry is preserved, and at low energies we get the Goldstone multiplet and a CS theory,
    \begin{equation}
    \Phi_G\,+\, SU(2)_{k_1-k_2-2}\quad\text{TQFT}.
\end{equation}
    \item When $k_1=k_2+1$, supersymmetry is broken, and we get in the IR
    \begin{equation}
    \phi_G\,+\,G_{\alpha}\,+\,U(1)_{2}\quad\text{TQFT},
    \end{equation}
    where $\phi_G$ is the Goldstone boson.
\end{itemize}

\subsubsection{Abelian vacua}
Finally, we look for Abelian vacua. There is still a solution given by
\begin{equation}
    \Phi=\left(
    \begin{matrix}
    v_3 && 0\\
    0 && 0
    \end{matrix}
    \right),\quad v_3=\frac{F_1+F_4}{m}.
\end{equation}
Introducing the critical value $\frac{g_1}{g_2}=\alpha$ such that $F_1(\alpha)+F_4(\alpha)=0$, we see that the solution exists either for $m>0,\,\tfrac{g_1}{g_2}>\alpha$, or for $m<0,\,\tfrac{g_1}{g_2}<\alpha$. At the point $\tfrac{g_1}{g_2}=\alpha$ the quantum superpotential again develops an asymptotic direction with zero energy.

We also note that, as follows from the definition of $F_4$, $\alpha<\left(\tfrac{k_2}{k_1}\right)^{1/4}$. This means that when we gradually move along the $m=0$ line from the region with $g_1\gg g_2$ to the region with $g_1\ll g_2$, we first see the flipping point for the Non-Abelian vacuum (where it moves from the positive mass region to the negative mass region), and then the flipping point for the Abelian vacuum.

To determine the IR physics of this vacuum, one has to reexamine the fermionic mass spectrum. It follows from \eqref{ThirdVacuumNeutralDiagonal}-\eqref{ThirdVacuumChargedMixingTwo} that upon passing by the point $\frac{g_1}{g_2}=\alpha$, the charged modes do not change mass signs (but one neutral Dirac mode does), thus the IR description is given by
\begin{equation}
    U(1)_{2(k_1-k_2)}\quad\text{TQFT}.
\end{equation}

Yet, this is not the end of the story. Indeed, when the point $\frac{g_1}{g_2}=\frac{k_2}{k_1}$ is passed by, a new solution for the equations $\partial_{|B|}\mathcal{W}=\partial_{\rho}\mathcal{W}=0$ is found. This can be seen in the following way. We note that $0\leq|B|\leq\rho$, and
\begin{subequations}
\begin{align*}
&\partial_{|B|}\mathcal{W}\,\vline_{|B|=\rho}>0,\\
&\partial_{|B|}\mathcal{W}\,\vline_{|B|=0}=0
\end{align*}
\end{subequations}
for any values of the parameters. But $\partial^2_{|B|}\mathcal{W}\propto F_3$ changes the sign exactly at the point $\frac{g_1}{g_2}=\frac{k_2}{k_1}$. In fact, when $\frac{g_1}{g_2}>\frac{k_2}{k_1}$, $\partial^2_{|B|}\mathcal{W}\,\vline_{|B|=0}>0$, and so it is possible that $B=0$ is the only zero of $\partial_{|B|}\mathcal{W}=0$: indeed, this is confirmed by the numerical study of the superpotential. On the other hand, when $\frac{g_1}{g_2}<\frac{k_2}{k_1}$, $\partial^2_{|B|}\mathcal{W}\,\vline_{|B|=0}<0$, and so there is at least one more solution with $B\neq 0$: the numerical study confirms that there is indeed only one such solution. This new vacuum still breaks the gauge group down to $U(1)$, but it also breaks the global $U(1)_B$. The IR physics is thus represented by
\begin{equation}\label{V4}
    \Phi_G+U(1)_{2(k_1-k_2)}\quad\text{TQFT}.
\end{equation}

\subsubsection{Phase Diagrams}
We now summarize the picture we suggest for the phase diagram, starting from the case $k_1,k_2>1$ and keeping in mind the relation in \eqref{alpharelations}. We recall that we started by determining the large mass phases, depicted by the purple and the green regions of Figure \ref{PhaseDiag4}. The next step was to understand the near-the-wall behaviour. When $g_1\gg g_2$, there is just one vacuum on the left from the wall, but two new vacua, the Non-Abelian and the Abelian ones, appear on the right from the wall (orange region of Figure \ref{PhaseDiag4}). While moving down along the wall, we encounter the first special point, after which the Non-Abelian vacuum is found on the left from the wall, while the Abelian vacuum is still on the right: this corresponds to the light blue and the yellow regions. In both these phases there is also a vacuum at the origin ($v_q$, where the $q$ stands for \textit{quantum}). We do not have a weak coupling limit that would allow the direct study of this vacuum, but we propose that its IR description is identical to the large negative mass vacuum, $v_1^-$, since it provides the correct Witten index, and automatically matches the UV 1-form symmetry 't Hooft anomaly.

When we decrease $\frac{g_1}{g_2}$ even further, the second special point is found. While passing it, we find that there is just one vacuum on the right from the wall (the green phase in Figure \ref{PhaseDiag4}), and three vacua on the left: a vacuum at the origin together with the Non-Abelian and the Abelian vacua discussed above (the grey phase in Figure \ref{PhaseDiag4}). The vacuum at the origin is now identified with the large mass vacuum.

If we go even further down the wall, the Abelian vacuum splits into two Abelian vacua ($v_3^-$ and $v_4$ in the brown region of Figure \ref{PhaseDiag4}). The $v_4$ vacuum was described above \eqref{V4}, and $v_3^-$ does not differ much from $v_3^+$: in fact, only the counter-terms for background fields associated to the global symmetry (e.g. the metric) are going to be different. 
 
\begin{figure}[h]
    \centering
    \includegraphics[width=0.6\textwidth]{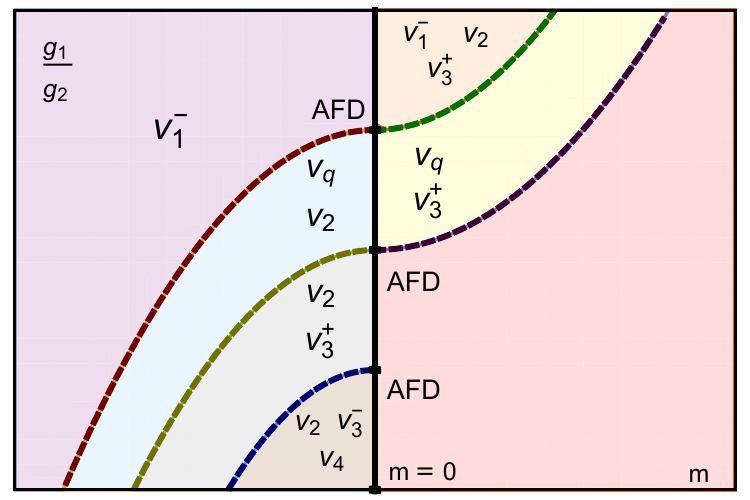}
    \caption{Phase diagram of the $SU(2)_{k_1}\times SU(2)_{-1}$ quivers with $k_1>1$. Dashed lines correspond to the second order phase transitions, while the solid line is the wall. The supersymmetric vacua in each phase are indicated.}
    \label{PhaseDiag5}
\end{figure}

The special case of $k_1>1$ and $k_2=1$, is pretty much similar, and the resulting phase diagram is depicted in Figure \ref{PhaseDiag5}. We note though that in this case the large positive mass phase (pink region) does not have any supersymmetric vacua, consequently, there are just two vacua in the grey phase and three vacua in the brown phase. It also implies that at the transition line between the yellow region and the grey region two supersymmetric vacua collide and, instead of producing a new supersymmetric vacuum, get lifted. In the other special case, when $k_2=k_1+1$, in the $v_2$ vacuum supersymmetry is broken and so the phase transition between the purple phase and the light blue phase is absent.
 
We conclude this section with the following observation. While sitting exactly at the wall and moving along it, we notice that at the flipping point for $v_3$ (the second special point) the Witten index jumps (in fact, for $k_2=1$ it vanishes below the flipping point, while being non-zero above). This is consistent with the fact that an asymptotically flat direction opens up at this point. Indeed, the transition can be arranged in the following way: at the special point one vacuum goes away to infinity, while another vacuum comes in from there. We thus conclude that exactly at the second flipping point there are no supersymmetric vacua, and the model exhibits the runaway behaviour (however, there might be meta-stable supersymmetry breaking vacua). The first flipping point does not show any special behaviour for the $SU(2)\times SU(2)$ quivers, but, as we will see below, there is an analogous phase transition at the first flipping point for the $SU(2)\times U(2)$ quivers. We thus expect the following picture: when we move down along the wall, at the first flipping point the supersymmetric vacuum goes away to infinity, and after this point the identical vacuum comes in from infinity. Again, at the special point the runaway behaviour takes place.

\section{Phase diagrams of $SU(2)\times U(2)$ models}\label{SUU}

In this section we discuss the phases of $SU(2)\times U(2)$ quiver theories, again with one bi-fundamental multiplet. Even though in principle one should recompute the effective superpotential for this case, we will appeal to a shortcut, and just assume that the vacuum structure (and in particular the symmetry breaking patterns) are the same as we have seen before. The main motivation for this assumption is that the Abelian factor inside U(2), for large CS level, does not modify the behaviour of the vacua.

\subsection{Chern-Simons levels of the same sign}\label{SUUCSLevelsSameSign}

We start by considering the models of the form
\begin{equation}
    \mathcal{N}=1\quad SU(2)_{k_1}\times U(2)_{k_2,k_3}
\end{equation}
with the coupling to bi-fundamental matter, and we will restrict ourselves with the case of $k_1\geq0$, $k_2>1$.\footnote{Consistency requires that $k_2=k_3+1\,\text{mod}\,2$, such that the resulting IR TQFTs satisfy the condition stated after \eqref{UConsistency}.}
\subsubsection{Large mass asymptotic phases}
As before, we can readily understand the large mass phases. When the mass is large and positive, we get the IR theory

\begin{equation}
SU(2)_{k_1}\times U(2)_{k_2,k_3+1}\quad\text{TQFT}
\end{equation}
with the index\footnote{In the case of $U(N)_{k_2,k_3}$ the Witten index reads 
\begin{equation*}
\text{WI}=\frac{(k_2+N-1)!\ k_3}{N!\ k_2!}    
\end{equation*}}
\begin{equation}
    \text{WI}_+=-\frac{(k_1+1)(k_2+1)|k_3+1|}{2}.
\end{equation}
When the mass is large and negative, few different cases can be discussed:
\begin{itemize}
    \item When $k_1>1$ and $k_2>1$, there is one supersymmetric vacuum whose IR theory is given by
    \begin{equation}
        SU(2)_{k_1-2}\times U(2)_{k_2-2,k_3-1}\quad\text{TQFT}.
    \end{equation}
    The Witten index is
    \begin{equation}
        \text{WI}_-=-\frac{(k_1-1)(k_2-1)|k_3-1|}{2}.
    \end{equation}
    \item When $k_1=1$, supersymmetry is spontaneously broken, and the IR theory is given by
    \begin{equation}
        G_{\alpha}+ U(1)_2 \times U(2)_{k_2-2,k_3-1}\quad\text{TQFT}.
    \end{equation}
    \item When $k_1=0$, we again see a supersymmetric vacuum hosting a CS theory,
    \begin{equation}
        U(2)_{k_2-2,k_3-1}\quad\text{TQFT},
    \end{equation}
    and the index is
    \begin{equation}
        \text{WI}_-=\frac{(k_2-1)|k_3-1|}{2}.
    \end{equation}
\end{itemize}
Following the familiar strategy, it is then useful to understand the dynamics near the wall, $m=0$, which we do now.

\subsubsection{Non-Abelian vacuum}
By assumption, there again exists a vacuum of the form
\begin{equation*}
    \Phi=\left(
    \begin{matrix}
    v_2 && 0 \\
    0 && v_2&
    \end{matrix}
    \right).
\end{equation*}
The gauge group is still broken to $SU(2)$ with the induced CS level $k_1+k_2$, but since the baryonic symmetry is now gauged, there are no Goldstone modes in the IR. In fact, the would-be Goldstone boson superpartner $\psi_G$ gets mixed with the $U(1)$ gaugino via the mass matrix

\begin{equation}
    \left(
    \begin{matrix}
    0 && -g_2 v_2\\
    -g_2 v_2 && -\kappa_3
    \end{matrix}
    \right).
\end{equation}

We assume that the rest of the fermionic spectrum is qualitatively the same, and thus the IR theory is given by
\begin{equation}
    SU(2)_{k_1+k_2-2}\quad\text{TQFT}.
\end{equation}
There are ten negative-mass Majorana modes, so the Witten index is
\begin{equation}
    \text{WI}_2=k_1+k_2-1.
\end{equation}
\subsubsection{Abelian vacuum}
In the same way we expect to find a vacuum of the form
\begin{equation*}
    \Phi=\left(
    \begin{matrix}
    v_3 && 0 \\
    0 && 0&
    \end{matrix}
    \right).
\end{equation*}
It breaks the gauge group to $U(1)\times U(1)$, and the induced CS levels are given by the matrix
\begin{equation}
   K= \left(
    \begin{matrix}
    2(k_1+k_2) && -k_2\\
    -k_2 && \frac{1}{2}(k_2+k_3)
    \end{matrix}
    \right).
\end{equation}
We can now use the fermionic charges and masses computed in Section \ref{SameCSLevelsThirdVac} to obtain the quantum corrections to the level matrix induced upon the integration out of the fermions:
\begin{equation}
    K_{IR}=\left(
    \begin{matrix}
    2(k_1+k_2) && -k_2\\
    -k_2 && \frac{1}{2}(k_2+k_3)+\frac{1}{2}
    \end{matrix}
    \right).
\end{equation}
The Witten index of this vacuum is given (up to a sign) by the number of lines in the corresponding Abelian CS theory,
\begin{equation}
    \text{WI}_3=\text{det}K_{IR}=-|(k_1+k_2)(k_2+k_3)+(k_1+k_2)-k_2^2|.
\end{equation}

\vspace{10pt}
\begin{center}
$ {\ast}\,{\ast}\,{\ast} $
\end{center}

The overall structure of the phase diagram is identical to the one depicted on Figures \eqref{PhaseDiag1},\eqref{PhaseDiag2}. We conjecture (following the pattern discussed in Section \ref{SameSignCSlevels}) that for $k_3\neq1$ at the intermediate (yellow) phase there is still the Abelian vacuum, as well as some other vacuum, resulting from the merging of the Non-Abelian vacuum and the vacuum at the origin. This \textit{quantum} vacuum is expected to support a TQFT or/and a non-linear sigma model with the Witten index fixed by the matching condition. When $k_3=1$, the vacuum structure happens to be quite different, and will be discussed in Section \ref{Dualities}.

\subsection{Chern-Simons levels of the opposite signs}\label{SUUCSlevelsofOppositeSigns}
Next we discuss the models of the form
\begin{equation}
    \mathcal{N}=1\quad SU(2)_{k_1}\times U(2)_{-k_2,-k_3}
\end{equation}
with, $k_1>k_2>1$.
\subsubsection{Large mass asymptotic phases}
When the matter mass is large and positive, we obtain
\begin{equation}
    SU(2)_{k_1}\times U(2)_{-k_2+2,-k_3+1}
\end{equation}
in the IR. There are three or four negative mass Majorana modes, depending on whether $k_3$ is positive or negative, so the index is
\begin{equation}
    \text{WI}_1=-\frac{(k_1+1)(k_2-1)(k_3-1)}{2}.
\end{equation}
When instead the mass is large and negative, we find a supersymmetric vacuum with
\begin{equation}
        SU(2)_{k_1-2}\times U(2)_{-k_2,-k_3-1}\quad\text{TQFT},
\end{equation}
\begin{equation}
        \text{WI}_1=\text{sgn}(k_3)\frac{(k_1-1)(k_2+1)|k_3+1|}{2}.
\end{equation}


\subsubsection{Non-Abelian vacuum}
Similarly to the $SU(2)\times SU(2)$ case, the non-Abelian vacuum is expected to exist on the right from the wall for $g_1\gg g_2$, and on the left from the wall for $g_1\ll g_2$, with a flipping point for some value of $\frac{g_1}{g_2}$. The gauge group is broken to $SU(2)_{k_1-k_2}$. The masses of fermions transforming in the adjoint representation don't change upon the crossing of the flipping point, there are always one of them with a positive mass and three with negative masses. On the contrary, one of the neutral Majorana fermions change the sign of its mass, such that there are seven negative-mass Majorana modes when $m>0$ and eight negative-mass Majorana modes when $m<0$. We therefore get in the IR
\begin{equation}
    SU(2)_{k_1-k_2-2}\quad\text{TQFT},\quad\text{WI}=-\text{sgn}(m)(k_1-k_2+1).
\end{equation}

\subsubsection{Abelian vacuum}
Finally, we suppose that there is an Abelian vacuum supporting the $U(1)\times U(1)$ CS theory. This vacuum is also expected to flip from one side of the wall to another at some value of $\frac{g_1}{g_2}$ (the second special point), and fermions charged under the unbroken gauge group do not flip the signs of their masses, and so the level matrix is given by
\begin{equation}
    K^+_{IR}=\left(
    \begin{matrix}
    2(k_1-k_2) && k_2\\
    k_2 && -\frac{1}{2}(k_2+k_3)+\frac{1}{2}
    \end{matrix}
    \right).
\end{equation}
on both sides of the wall. Some neutral modes though flip their masses, so that the Witten index is negative for $m>0$ and positive for $m<0$.

Decreasing the ratio $\frac{g_1}{g_2}$ even further, we expect to face the third special point where a new Abelian vacuum with $B\neq 0$ appears ($v_4$ in the brown region of Figure \ref{PhaseDiag4}). This vacuum preserves just one Abelian factor, and supports
\begin{equation}
    U(1)_{2(k_1-k_2)}\quad\text{TQFT}
\end{equation}
in the IR. The Abelian vacuum discussed above also undergoes some changes when the third special point is passed. Namely, one of the fermions charged under the second $U(1)$ gets a negative mass, which leads to the corrected level matrix:
\begin{equation}
    K^-_{IR}=\left(
    \begin{matrix}
    2(k_1-k_2) && k_2\\
    k_2 && -\frac{1}{2}(k_2+k_3)-\frac{1}{2}
    \end{matrix}
    \right).
\end{equation}
The Abelian vacuum with the $U(1)\times U(1)$ gauge group and the level matrix given above is denoted by $v_3^-$ in Figure \ref{PhaseDiag4}.

\begin{center}
$ {\ast}\,{\ast}\,{\ast} $
\end{center}
The overall phase diagram looks quite similar to what is seen in Figures \ref{PhaseDiag4} and \ref{PhaseDiag5}, despite it is now harder to guess the vacuum at the origin in yellow and purple phases. We conjecture that for $k_2=2, k_3=1$ it is given by the Abelian CS theory
\begin{equation}
    U(1)_{2k_1}.
\end{equation}
This conjecture is motivated by a duality discussed in the next section. Slightly more generally, for $k_3=1$ (in which case the Witten index of the large mass phase vanishes) it is natural to expect that this vacuum supports the same TQFT in the IR, as does the Abelian vacuum also existing in this phase, namely
\begin{equation}
    \left[U(1)\times U(1)\right]_K\quad\text{TQFT},
\end{equation}
\begin{equation}
    K=\left(
    \begin{matrix}
    2(k_1-k_2) && k_2\\
    k_2 && -\frac{1}{2}k_2
    \end{matrix}
    \right),
\end{equation}
but with the opposite value of the Witten Index. The two proposal are consistent, if for $k_2=2$, the theory $\left[U(1)\times U(1)\right]_K$ is dual to $U(1)_{2k_1}$. This is indeed the case, as can be seen by conjugating the matrix $K$ with the unimodular matrix\footnote{We are grateful to the referee for pointing this out to us.}
\begin{equation}
    U=\left(
    \begin{matrix}
    1 & 2\\
    0 & 1
    \end{matrix}
    \right),
\end{equation}
and obtaining
\begin{equation}
    U\,K\,U^T=\left(
    \begin{matrix}
    2k_1 & 0\\
    0 & -\tfrac{1}{2}k_2
    \end{matrix}
    \right).
\end{equation}
For $k_2=2$, this last representation  corresponds to $U(1)_{2k_2}$ as a spin TQFT.

The last comment concerns the dynamics at the special points. Similarly to the discussion at the end of Section \ref{SUSUCSlevelsofOppositeSigns}, we observe two phase transitions, at the first and at the second special points. As before, they are organized by first sending a supersymmetric vacuum to infinity, and then receiving a new supersymmetric vacuum, generically with a different TQFT and Witten index, from infinity, with the runaway behaviour at the transition point.

\section{Dualities}\label{Dualities}
The discussion of the previous two sections demonstrated that a generic 3d $\mathcal{N}=1$ quiver gauge theory has multiple second-order phase transitions with associated IR fixed points. In this section we will provide few conjectures stating that certain CFTs that appear as IR limits of different quiver theories may in fact be the same: this is the statement of the IR duality.
Some of such dualities were already used above to guess certain aspects of the phase diagrams (namely, the vacuum structures in the intermediate phases).

\subsection{Dualities between $SU(2)\times SU(2)$ and $SU(2)\times U(2)$}

\begin{figure}
  \centering
  \subfigure[]{\label{PhaseDiag7}\includegraphics[width=0.45\textwidth]{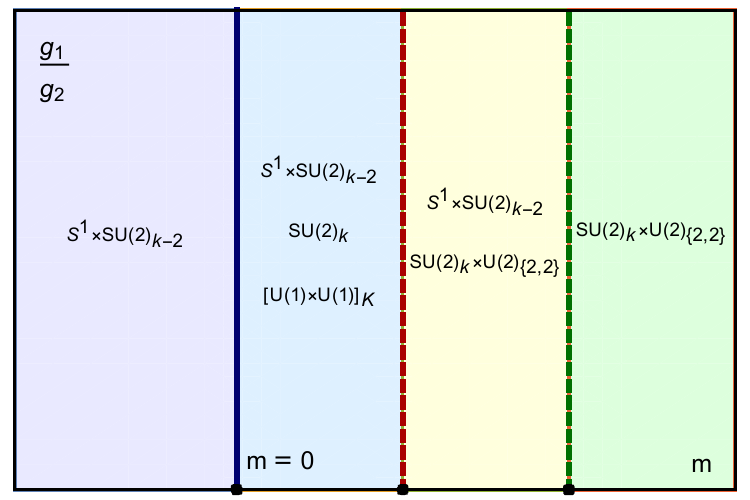}}\qquad
  \subfigure[]{\label{PhaseDiag8}\includegraphics[width=0.45\textwidth]{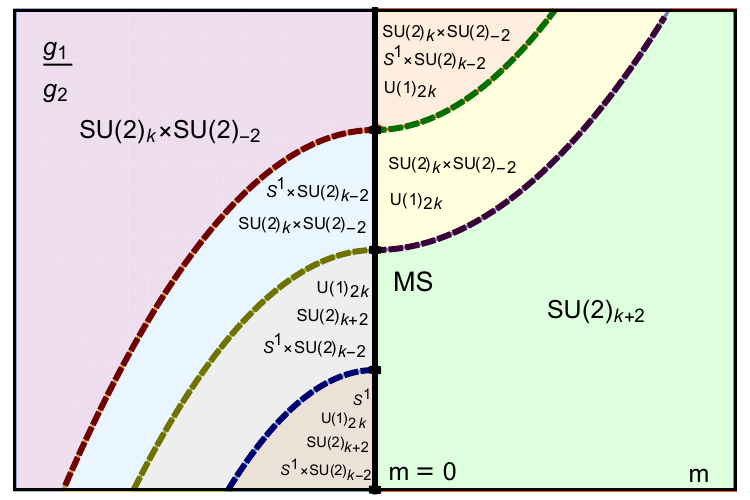}}
\caption{Phase diagrams for the $SU(2)_{k}\times U(2)_{2,1}$ quiver (a) and for the $SU(2)_{k+2}\times SU(2)_{-2}$ quiver (b).}
\label{Duality1}
\end{figure}

The first pair of theories we consider is $SU(2)_{k}\times U(2)_{2,1}$ and $SU(2)_{k+2}\times SU(2)_{-2}$ quivers with $k>0$; the corresponding phase diagrams are shown in Figure \ref{Duality1}, and the yellow phase of \ref{PhaseDiag7} as well as the yellow and light blue phases of \ref{PhaseDiag8} are conjectures. We observe, using the level-rank duality
\begin{equation}\label{LRduality}
    SU(2)_{-2}\longleftrightarrow U(2)_{2,2},
\end{equation}
that the transition in Figure \ref{PhaseDiag7} between the yellow and the green phases is identical to the transition in Figure \ref{PhaseDiag8} between the light blue phase and the purple phase, with the phases on both sides of the transition given by
\begin{equation}
    \mathbb S^1\times SU(2)_{k-2} + SU(2)_k\times SU(2)_{-2}\longrightarrow SU(2)_k\times SU(2)_{-2}.
\end{equation}

While making the conjecture about the vacua in the yellow phase of Figure \ref{PhaseDiag7}, we assumed that at the transition point $m=m_*$ the non-Abelian vacuum merges with the Abelian one, while the vacuum at the origin stays apart. This is in contrast with what was assumed in Sections \ref{SameSignCSlevels} and \ref{SUUCSLevelsSameSign}. The difference comes from the fact that here a new "branch" of vacua, parametrized by the dual photon, emerges.
 \begin{wrapfigure}{r}{0.3\textwidth}
 \includegraphics[width=0.45\textwidth]{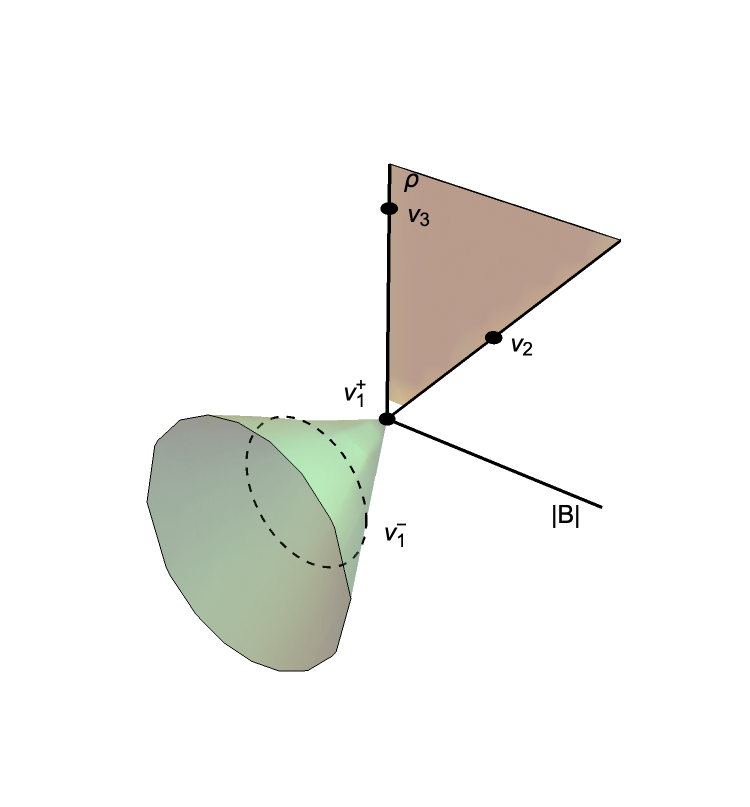}
 \end{wrapfigure}
The space of possible vacua can be then visualized as the two-dimensional space parametrized by the scalar vevs, together with a cone attached at the origin (see the Figure on the right). The angular direction of the cone is the dual photon, and the radial direction gives the radius of the circle (which is not a dynamical field, but rather a function of the parameters). It is then possible that first the non-Abelian and the Abelian vacua meet at the origin, and then the dual photon radius (as a function of $m$) shrinks to zero, and the second phase transition happens.  
 
\begin{figure}
  \centering
  \subfigure[]{\label{PhaseDiag9}\includegraphics[width=0.45\textwidth]{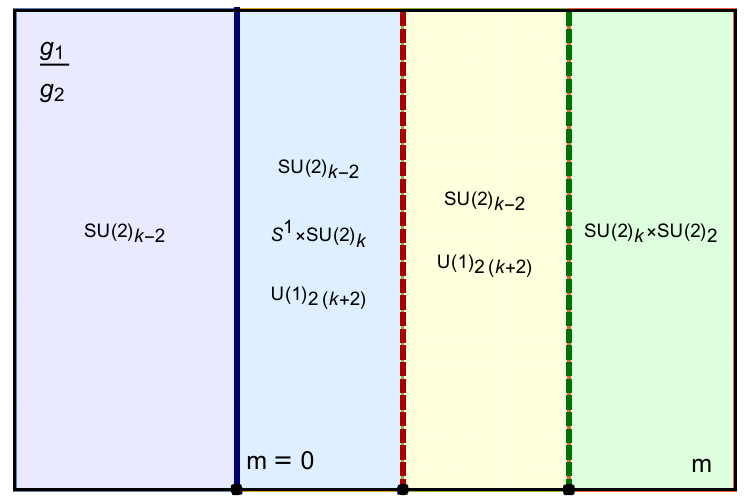}}\qquad
  \subfigure[]{\label{PhaseDiag10}\includegraphics[width=0.45\textwidth]{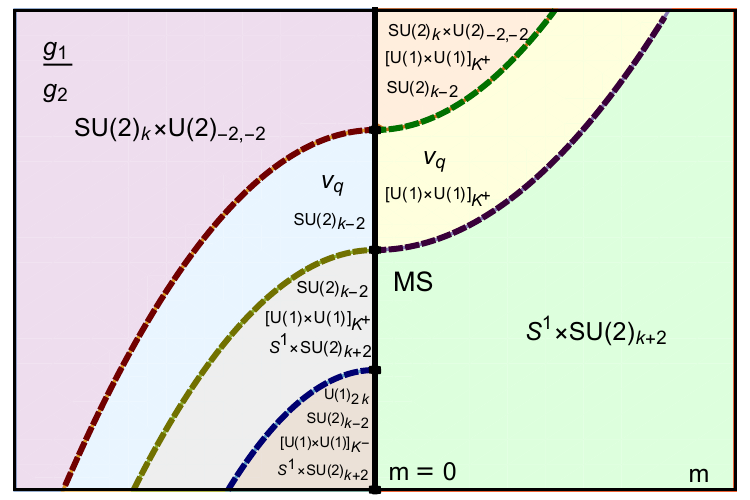}}
\caption{Phase diagrams for the $SU(2)_{k}\times SU(2)_2$ quiver (a) and for the $SU(2)_{k+2}\times U(2)_{-2,-1}$ quiver (b).}
\label{Duality2}
\end{figure}

The second pair is  $SU(2)_{k}\times SU(2)_{2}$ and $SU(2)_{k+2}\times U(2)_{-2,-1}$ quivers with $k>0$; the corresponding phase diagrams are presented in Figure \ref{Duality2}, and the yellow phase of Figure \ref{PhaseDiag9} is a conjecture. We propose that the "quantum" vacuum $v_q$ in Figure \ref{PhaseDiag10} is given by
\begin{equation}
    U(1)_{2(k+2)}\quad\text{TQFT}.
\end{equation}

Using again the level-rank duality \eqref{LRduality}, we observe that the transition in Figure \ref{PhaseDiag9} between the yellow and the green phases is identical to the transition in Figure \ref{PhaseDiag10} between the light blue phase and the purple phase,
\begin{equation}
    SU(2)_{k-2} + U(1)_{2(k+2)}\longrightarrow SU(2)_k\times SU(2)_{2}.
\end{equation}

The two dualities we have just described can be obtained from the duality \eqref{ParentDuality} by gauging the flavour $SU(2)$ (sub)groups on both sides.\footnote{One may wonder why the $SU(2)$ gauge fields used to gauge the common $SU(2)$ subgroup on both sides of the duality \eqref{ParentDuality} have different CS levels on the two sides of the duality claimed above. The reason for this is that the duality \eqref{ParentDuality} implies different contact terms for various background fields.}

\subsection{Duality between $SU(2)\times SU(2)$ quiver and adjoint QCD}

\begin{figure}
  \centering
  \subfigure[]{\label{PhaseDiag11}\includegraphics[width=0.45\textwidth]{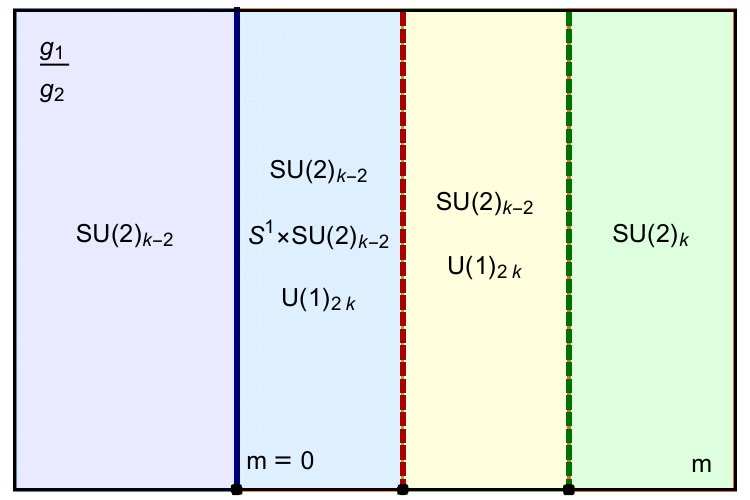}}\qquad
  \subfigure[]{\label{PhaseDiag12}\includegraphics[width=0.45\textwidth]{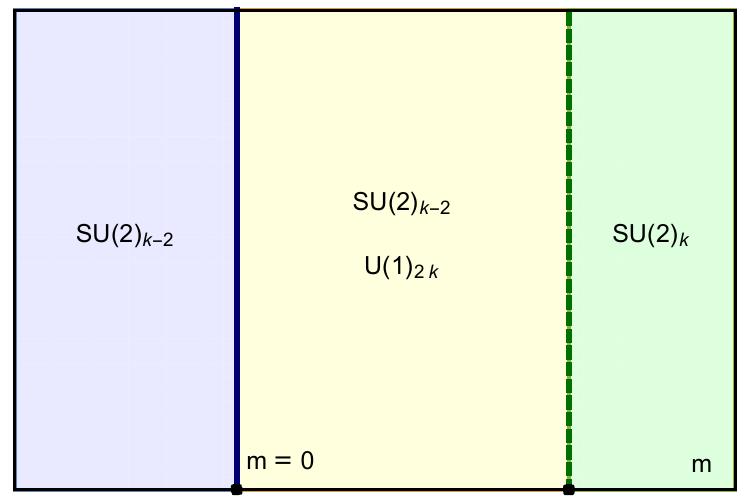}}
\caption{Phase diagram for the $SU(2)_k\times SU(2)_0$ quiver (a) and for the $SU(2)_k$ adjoint QCD (b).}
\label{Duality3}
\end{figure}

The first model considered here is the $SU(2)_k\times SU(2)_0$ quiver, discussed in section \ref{SameCSLevelsThirdVac}. The phase diagram can be found in Figure \ref{PhaseDiag11}, where the form of the yellow phase is a conjecture. The second model is the $SU(2)_k$ adjoint QCD: the corresponding phase diagram was reviewed in Section \ref{Review}, and is depicted on Figure \ref{PhaseDiag12}.

Evidently, the phase transitions between the yellow phases and the green phases are identical, and this hints towards the possibility of the duality:
\begin{equation}
    SU(2)_k\times SU(2)_0\quad\text{with a bi-fundamental}\quad\longleftrightarrow\quad SU(2)_k\quad\text{with an adjoint}.
\end{equation}
This duality if correct has a quite clear meaning. Assuming that the $SU(2)_0$ node of the quiver confines, we can describe the low-energy physics in terms of the bilinear
\begin{equation}
    X=\Phi\Phi^{\dagger},
\end{equation}
which indeed transforms in the adjoint representation of $SU(2)_k$. There is one point in this picture that may seem disturbing. The quiver theory possesses the baryonic symmetry $U(1)_B$, and there are charged operators $B=\det\Phi$.  Neither the symmetry, nor would-be dual operators appear on the QCD side. This issue can be resolved in two ways: either the quiver theory baryons happen to be massive, and do not appear in the IR fixed point, or they are actually massless at the CFT point, but decouple from the rest. In the latter case the QCD side should be supplemented by a decoupled free complex multiplet.

\section{Time reversal invariant models}\label{TimeReversalInvariantModels}

We have already mentioned that $3d$ $\mathcal{N}=1$ theories with time reversal invariance have a beautiful property: their superpotentials admit only corrections odd under the action of $T$-transformation \cite{Gaiotto:2018yjh}. It significantly restricts the possible form of the effective superpotential, and sometimes superpotential even turns out to be fully protected.

Examples of $T$-invariant theories can be found also among the quiver theories. For example, a two-node quiver with opposite CS levels,  
\begin{equation}
    \mathcal{N}=1\quad SU(2)_{k}\times SU(2)_{-k}\,+\,\text{a bi-fundamental},
\end{equation}
enjoys this property at the point $g_1=g_2$, $m=0$\footnote{To be more precise, $T$-transformation must be augmented by the exchange of two gauge group factors.}. It is easy to see that there are no parity odd terms that could be written in the effective superpotential, implying that we have an example with full protection at hands. In fact, one can check that the 1-loop superpotential computed in Section \ref{Superpotential} vanishes at this point. It follows that the theory has a moduli space of vacua, which coincides with the classical one,
\begin{equation}
  \mathcal M = \mathbb S^1\,\times\,\mathbb{R}^2\,/\,S_2.
\end{equation}
At the origin of the moduli space we expect to find a SCFT. At a point away from the origin the IR physics is described by three real massless moduli without any topological sector.

We can then deform the theory from the $T$-invariant point by turning on the mass term or changing the ratio $\frac{g_1}{g_2}$, and study the resulting IR phases. The large mass phases for $k>1$ are supersymmetric and are given by
\begin{equation}
SU(2)_{k}\times SU(2)_{-k+2}\quad\text{TQFT}
\end{equation}
for large positive masses and 
\begin{equation}
SU(2)_{k-2}\times SU(2)_{-k}\quad\text{TQFT}
\end{equation}
for large negative masses. When $k=1$, supersymmetry is spontaneously broken, and the IR description is
\begin{equation}
    G_{\alpha}+U(1)_2\times U(1)_2\quad\text{TQFT}
\end{equation}
for large positive masses, and the same for large positive masses, where the level-rank duality \eqref{LRduality2} has been used.
\begin{figure}[h]
    \centering
    \includegraphics[width=0.6\textwidth]{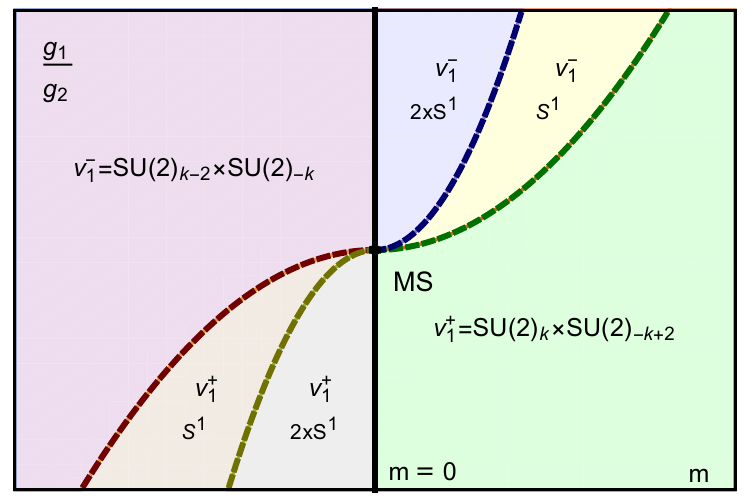}
    \caption{Structure of the phase diagram for the $SU(2)_k\times SU(2)_{-k}$ quiver. Dashed lines correspond to the second order phase transitions, while the solid line is the wall. The supersymmetric vacua in each phase are indicated.}
    \label{PhaseDiagT}
\end{figure}

We can also study behaviour near the wall and find that for $\frac{g_1}{g_2}>1$ two new vacua appear when $m>0$, while  for $\frac{g_1}{g_2}<1$  we find them when $m<0$. These are would-be Abelian and non-Abelian vacua familiar from above, despite the fact that they do not support any topological degrees of freedom. Still, they are not trivial and host $S^1$ Goldstone bosons: Abelian vacuum breaks spontaneously magnetic symmetry of the preserved $U(1)$ gauge group, while non-Abelian vacuum breaks the baryonic symmetry. The phase diagram is depicted in Figure \ref{PhaseDiagT}.

\section{Outlook}\label{Outlook}

In this paper we studied the IR behaviour of certain $3d$ $\mathcal{N}=1$ quiver theories. Already the simplest possible setup of a two-node quiver with $SU(2)$ gauge groups and one bi-fundamental multiplet reveals quite rich and diverse physical pictures. We find the characteristic features of theories with two supercharges observed previously in the literature: walls in the parameter space at which the Witten index jumps, multiple phases with second order phase transitions between them, vacua with spontaneously broken supersymmetry, which can be either stable or meta-stable. Especially interesting phase diagrams are found in theories with CS levels of different signs (Sections \ref{SUSUCSlevelsofOppositeSigns}, \ref{SUUCSlevelsofOppositeSigns}). The study of phase diagrams was facilitated by some duality considerations: conjecturing certain dualities, one sometimes could gain more understanding of the intermediate "quantum" phases, which are not continuously connected to any weakly coupled limit.

Two of the dualities considered in the paper can be related to previously conjectured dualities for SQCD-like theories, and obtained from them by gauging flavour symmetries on both sides. This tool for generating new dualities is well-known for theories with greater amount of supersymmetry, but our results suggest that it is also applicable in the realm of minimally supersymmetric three-dimensional theories. The third duality we discuss involves the confinement of a node, and corresponds to the situation for which at low energies the physics can be described in terms of bilinears (or more generally multi-linears) with respect to the original matter field, and gauge-neutral with respect to a given node.

Many results of this paper offer directions for generalizations. The most obvious one is to consider quivers with higher-rank gauge groups, as $SU(N)$ or $U(N)$. The overall structure of the phase diagrams is expected to be more complicated in these cases but nevertheless one may hope to understand the dynamics near the wall (even though the 1-loop superpotential is going to have a more complicated structure). More general dualities of \cite{Choi:2018ohn} can then be used, together with the node-dualization technique, to conjecture new dualities between quivers. In this case, a more detailed analysis is required in order to establish to which of the multiple SCFTs the duality applies.

Another interesting generalization is to take several bi-fundamental multiplets. The resulting vacua will generically break flavour symmetries, therefore the IR description will be given in terms of non-linear sigma models, in addition to the topological sectors, similar to those already seen above.
 
We have already mentioned in the introduction that simple quiver theories, like the ones that we have been studying in this work, are prototypical examples of more complicated models arising as world-volume theories on M2 branes put on certain backgrounds \cite{Forcella:2009jj}. Since classically these theories have moduli spaces, it is natural to think that they appear exactly on the walls inside their phase diagrams. Borrowing some lessons from the results we found above, we expect that at the quantum level different possibilities can arise: the vacuum may be supersymmetric and host some topological sector, or it may break supersymmetry, or show a runaway behaviour with an asymptotically flat direction. On the other hand, it is less likely that the IR limit of the M2 brane world-volume theory is given by a SCFT, unless the theory is $T$-invariant. It is extremely interesting to investigate which of these possibilities are realized in various string/M-theory configurations.

To conclude, there remain many exciting questions regarding the dynamics of three-dimensional quivers with minimal supersymmetry, and we hope to report on the progress in some of them in the future.   

\vskip 50pt

\section*{Acknowledgements}
We thank Zohar Komargodski, Joseph Minahan, and Adar Sharon for valuable discussions. The work of VB is supported by the European Research Council (ERC) under the European Union’s Horizon 2020 research and innovation programme (grant agreement No. 851931). The work of NG is supported by the “Gauge Theories, Strings, Supergravity” (GSS) research project.

\newpage

\appendix

\section{Group Theory Conventions}
The gauge group we consider in this paper is of the form $G=G_1\times G_2$. In the main text $G_{1,2}$ are chosen to be $SU(2)$ so we restrict to this one. \\
The $\mathfrak{su}(2)$ Lie algebra is defined through the usual relation
\begin{equation}
[\tau^A,\tau^B]=i\epsilon^{ABC}\tau^C    
\end{equation}
where $A,B,C$ are adjoint indices. We choose $\tau^A\equiv\frac{1}{2}\sigma^A$, where $\sigma^A$ are the Pauli matrices, so that the canonical normalization 
\begin{equation}
\text{Tr}\left(\tau^A \tau^B\right)=\frac{1}{2}\delta^{AB}    
\end{equation}
holds. We make also use of the following notation
\begin{equation}
\tau^{(A}\tau^{B)}\equiv\frac{1}{2}\{\tau^A, \tau^B\},\qquad \tau^{[A}\tau^{B]}\equiv\frac{1}{2}[\tau^A, \tau^B]
\end{equation}
Since we are considering a quiver gauge theory with the matter sitting in the (anti-)bifundamental representation of $G_1\times G_2$, namely $(\bar R, R)$ and $(R,\bar R)$ respectively, we recall the action of $G$ on them. The gauge group transformations $U\in G_1$ and $V\in G_2$ will then act in the following way
\begin{equation}
(\Phi')_i^{\ \hat j}=(U\Phi V^\dagger)_i^{\ \hat j},\qquad 
(\bar \Phi')_{\hat i }^{\ j}=(V\bar\Phi U^\dagger)_{\hat i }^{\ j}
\end{equation}
and at the level of algebra we have
\begin{equation}
 \delta^{(A)}\Phi_i^{\ \hat j}=i\left[g_1 (T^{(A)})_i^{\ k}\Phi_k^{\ \hat j}-g_2\Phi_i^{\ \hat k}(K^{(A)})_{\hat k}^{\ \hat j}\right ], \qquad  \delta^{(A)}\bar \Phi_{\hat i}^{\ j}=i\left[g_2 (K^{(A)})_{\hat i}^{\ \hat k}\bar \Phi_{\hat k}^{\ j}-g_1 \Phi_{\hat i}^{\ k}(T^{(A)})_{k}^{\ j}\right ]
\end{equation}
where $g_{1,2}$ are the two gauge couplings for $G_{1,2}$ and $T^{(A)}$, $K^{(A)}$ are generators of the $\mathfrak{g}_1, \mathfrak{g}_2$ Lie algebras respectively.

\section{Superspace Conventions}
The conventions we used in the main text refer to \cite{Gates:1983nr, Choi:2018ohn}. \\
We raise and lower spinor indices through the use of the antisymmetric matrix $C_{\alpha\beta}$ as follows
\begin{equation}
\psi^\alpha=C^{\alpha\beta}\psi_\beta,\quad \psi_\alpha=\psi^\beta C_{\beta\alpha}=-C_{\alpha\beta}\psi^\beta
\end{equation}
with $\psi^2\equiv\frac{1}{2}\psi^\alpha\psi_\alpha$.
The graded commutation relation for derivatives is
\begin{equation}
\{D_\alpha, D_\beta\}=2i\partial_{\alpha\beta}    
\end{equation}
where $\partial_{\alpha\beta}$ is the ordinary spacetime derivative. Derivatives also satisfy the following useful identities
\begin{equation}\label{identities}
\begin{aligned}
&\qquad\qquad\partial^{\alpha\gamma}\partial_{\beta\gamma}=\delta^\alpha_\beta\Box,\quad D_\alpha D_\beta=i\partial_{\alpha\beta}+C_{\alpha\beta}D^2\\
&D_\alpha D_\beta D^\alpha=0, \quad D^2 D_\alpha= -D_\alpha D^2=i\partial_{\alpha\beta}D^\beta, \quad (D^2)^2=\Box
\end{aligned}
\end{equation}
\subsection{$\mathcal{N}=1$ Gauge Theories}
The most generic kinetic terms for three-dimensional supersymmetric gauge theories are Chern-Simons and Yang-Mills ones. In the $\mathcal{N}=1$ notation, the Lagrangian for such terms reads
\begin{equation}
\mathcal{L}_{\text{CS-YM}}=-\frac{k}{4\pi}\Tr(2i\Gamma^\alpha\partial_{\alpha\beta}\Gamma^\beta+\Gamma^\alpha D_\alpha D^\beta\Gamma_\beta)+\frac{1}{2g^2}\Tr(\Gamma^\alpha\Box\Gamma_\alpha-i\Gamma^\alpha\partial_{\alpha\beta}D^2\Gamma^\beta)    
\end{equation}
which, with some effort, can be recasted in the following gauge-fixed form 
\begin{equation}
\begin{aligned}
\mathcal{L}_{\text{CS-YM}}^{\text{gf}}=  &-\frac{k}{4\pi}\Tr(2i\Gamma^\alpha\partial_{\alpha\beta}\Gamma^\beta+\left(1-\frac{1}{\beta}\right)\Gamma^\alpha D_\alpha D^\beta\Gamma_\beta)\\
&+\frac{1}{2g^2}\Tr(\left(1+\frac{1}{\alpha}\right)\Gamma^\alpha\Box\Gamma_\alpha-i\left(1-\frac{1}{\alpha}\right)\Gamma^\alpha\partial_{\alpha\beta}D^2\Gamma^\beta) \end{aligned}
\end{equation}
By taking the Landau gauge-fixing limit ($\alpha,\beta\rightarrow 0$) one can obtain the final form for the CS-YM gauge propagator, which reads
\begin{equation}\label{gaugeprop}
\Delta_\alpha^{\ \beta}=g^2\frac{\delta_\alpha^\beta(\kappa D^2+p^2)+(\kappa-D^2)p_\alpha^{\ \beta}}{p^2(\kappa^2+p^2)},\qquad \kappa=\frac{k g^2}{2\pi}  
\end{equation}
Matter can be also coupled through the following action 
\begin{equation}
S_\text{matter}=-\frac{1}{2}\int d^3x d^2\theta\ (\nabla^\alpha\bar\Phi)(\nabla_\alpha\Phi) 
\end{equation}
where in our setup, the covariant derivatives take the following explicit form
\begin{align}
    \nabla_{\alpha}{\Phi_i}^{\hat j} &=D_{\alpha}{\Phi_i}^{\hat j}-ig_1\ \Gamma^A_\alpha{(T^A)_i}^{k}{\Phi_{k}}^{\hat j}+ig_2\ \hat\Gamma^M_\alpha\ {\Phi_i}^{\hat k}{(K^M)_{\hat k}}^{\hat j}\\
    \nabla_{\alpha}{{\bar\Phi}_{\hat j}}^{\ i} &=D_{\alpha}{\bar\Phi_{\hat j}}^{\ i}-ig_2\ \hat\Gamma^M_\alpha{(K^M)_{\hat j}}^{\hat k}{\bar\Phi_{\hat k}}^{\ i}+ig_1\ \Gamma^A_\alpha\ {\bar \Phi_{\hat j}}^{\ k}{(T^A)_{k}}^{i}
\end{align}

\section{Effective superpotential at $k_1=-k_2$}\label{Crocodiles}

When the CS levels are equal by the absolute value but opposite, one can compute derivatives of the superpotential explicitly. Introducing $k_1=-k_2=k$, $\kappa_1=\frac{k g_1^2}{2\pi}$, $\kappa_2=\frac{k g_2^2}{2\pi}$, we get for $B\rightarrow\rho$:
\vspace{10pt}
\begin{subequations}
\begin{align}
    &\partial_{\rho}\mathcal{W}_{\text{1-loop}}=-\frac{(g_1^2+g_2^2)(\kappa_1-\kappa_2)\left[(9g_1^4 + 22 g_1^2 g_2^2 + 9 g_2^4) \rho+6 (g_1^2 + g_2^2) (\kappa_1^2 + \kappa_1 \kappa_2 + \kappa_2^2)\right]}{16\pi(2\kappa_1 \kappa_2+\rho(g_1^2+g_2^2))\sqrt{(\kappa_1+\kappa_2)^2+2\rho(g_1^2+g_2^2)}}\nonumber\\
    &\qquad\qquad\qquad\qquad\qquad\overset{\rho\rightarrow\infty}{\longrightarrow}-\frac{(\kappa_1-\kappa_2)(9g_1^4 + 22 g_1^2 g_2^2 + 9 g_2^4) }{16\pi\sqrt{2}\sqrt{(g_1^2+g_2^2)}\rho^{1/2}},\\
    &\partial_{|B|}\mathcal{W}_{\text{1-loop}}=\frac{g_1^2 g_2^2 (\kappa_1-\kappa_2)\rho}{4\pi(2\kappa_1\kappa_2+\rho(g_1^2+g_2^2))\sqrt{(\kappa_1+\kappa_2)^2+2\rho(g_1^2+g_2^2)}}\nonumber\\
    &\qquad\qquad\qquad\qquad\qquad\overset{\rho\rightarrow\infty}{\longrightarrow}\frac{g_1^2 g_2^2 (\kappa_1-\kappa_2)}{4\sqrt{2}\pi(g_1^2+g_2^2)^{3/2}\rho^{1/2}},
\end{align}
\end{subequations}
and in the limit $B\rightarrow0$ the result is
\begin{subequations}
\begin{align}
    &\partial_{\rho}\mathcal{W}_{\text{1-loop}}=-\frac{\kappa_1-\kappa_2}{8\pi(g_1^2-g_2^2)}\left[\frac{g_1^4}{\sqrt{\kappa_1^2+2g_1^2\rho}}-\frac{g_2^4}{\sqrt{\kappa_2^2+2g_2^2\rho}} \right]-\nonumber\\
    &\quad-\frac{(g_1^2+g_2^2)(\kappa_1-\kappa_2)}{16\pi\sqrt{(\kappa_1+\kappa_2)^2+2\rho (g_1^2+g_2^2)}}\nonumber\\
    &\qquad\qquad\qquad\overset{\rho\rightarrow\infty}{\longrightarrow}-\frac{(\kappa_1-\kappa_2)(g_1^3-g_2^3)}{8\sqrt{2}\pi(g_1^2-g_2^2)\rho^{1/2}}-\frac{(\kappa_1-\kappa_2)\sqrt{g_1^2+g_2^2}}{16\sqrt{2}\pi\rho^{1/2}},\\
    &\partial_{|B|}\mathcal{W}_{\text{1-loop}}=\frac{g_1^2g_2^2(\kappa_1-\kappa_2)B}{4\pi(g_1^2-g_2^2)D}\left[\frac{g_1^4+3g_1^2g_2^2}{\sqrt{\kappa_1^2+2g_1^2\rho}}-\frac{g_2^4+3g_1^2g_2^2}{\sqrt{\kappa_2^2+2g_2^2\rho}}\right]\nonumber\\
     &\qquad\qquad\qquad\overset{\rho\rightarrow\infty}{\longrightarrow}\frac{g_1g_2(\kappa_1 g_2-\kappa_2 g_2)}{4\sqrt{2}\pi(g_1+g_2)^2}.
\end{align}
\end{subequations}
where
\begin{subequations}
\begin{align}
    &D=(g_1^2-g_2^2)^2\rho-4(\kappa_1^2g_2^2+\kappa_2^2g_1^2).
\end{align}
\end{subequations}

\bibliographystyle{JHEP}
\bibliography{biblio}

\end{document}